\documentclass[aps,tightenlines,twocolumn,prd,nofootinbib,
superscriptaddress,showpacs,preprintnumbers,eqsecnum,floatfix]{revtex4-1}
\def\bea{\begin{eqnarray}}
\def\be{\begin{equation}}
\def\ee{\end{equation}}
\def\eea{\end{eqnarray}}
\def\bal{\begin{align}}
\def\eal{\end{align}}

\usepackage[usenames,dvipsnames]{color}
\usepackage{graphics}
\usepackage{graphicx}
\usepackage{amsmath}
\usepackage{amssymb}
\usepackage{bbold}
\usepackage{bm}
\usepackage{slashed}
\usepackage{float}
\usepackage{psfrag}
\usepackage{footnote} 
\usepackage{xcolor}
\usepackage{textcomp}
\usepackage{cancel}
\begin{document}
 \preprint{CFTP/18-014} 
 \preprint{JLAB-THY-18-2838}

\title{Quark mass function from a one-gluon-exchange-type interaction in Minkowski space  }

 

 \author{Elmar P. Biernat}
\affiliation{
CFTP, Instituto Superior T\'ecnico, Universidade de Lisboa, Avenida Rovisco Pais 1, 
1049 Lisboa, Portugal}
 \email{elmar.biernat@tecnico.ulisboa.pt}

\author{Franz Gross }
 \affiliation{ Thomas Jefferson National Accelerator Facility (JLab), Newport News, VA 23606, USA}
 \affiliation{College of William and Mary, Williamsburg, Virginia 23188,
USA}

 \author{M. T Pe\~na}
\affiliation{CFTP, Departamento de F\'isica, Instituto Superior T\'ecnico, Universidade de Lisboa, Avenida Rovisco Pais 1, 
1049 Lisboa, Portugal}

\author{Alfred Stadler}
\affiliation{Departamento de F\'isica, Universidade de \'Evora, 7000-671 \'Evora, Portugal}
\affiliation{CFTP, Departamento de F\'isica, Instituto Superior T\'ecnico, Universidade de Lisboa, Avenida Rovisco Pais 1, 
1049 Lisboa, Portugal}

\author{Sofia Leit\~ao}
\affiliation{
CFTP, Instituto Superior T\'ecnico, Universidade de Lisboa, Avenida Rovisco Pais 1, 
1049 Lisboa, Portugal}
\date{\today}
 \begin{abstract}
 We present first results for the quark mass function in Minkowski space in both the spacelike and timelike regions calculated from the same quark-antiquark interaction kernel used in the latest meson calculations using the Gross equation. This kernel consists of a Lorentz vector effective one-gluon-exchange-type interaction, a vector constant, and a mixed scalar-pseudoscalar covariant linear confining interaction that does not contribute to the mass function. We analyze the gauge dependence of our results, prove the gauge independence of the constituent quark mass and mass gap equation, and identify the Yennie gauge as the appropriate gauge to be used in CST calculations. We compare our results in the spacelike region to lattice QCD data and find good agreement.
 \end{abstract}


\maketitle



  \section{Introduction}

The highly non-perturbative nature of QCD in the low-energy regime makes the theoretical description of strong-interaction phenomena, such as confinement and dynamical chiral symmetry breaking (D$\chi$SB), very difficult. Nevertheless, many different approaches have been used to calculate the properties of QCD bound states and their reactions, such as lattice QCD~\cite{Aoki2017,Fodor:2012gf,PhysRevD.84.074508,Colangelo2011,Gattringer:2010zz,Greensite:2003bk,PhysRevD.64.054506,PhysRevD.12.2060,Bowman:2005vx,Oliveira:2018lln}, Bethe-Salpeter/Dyson-Schwinger (BSDS) equations~\cite{Eichmann:2016bf,PhysRevC.79.012202,Rojas2013,Maas:2011se,Binosi:2009qm,PhysRevD.75.087701,0954-3899-32-8-R02,Maris:2003vk,ALKOFER2001281,Tandy:1997qf,ROBERTS1994477}, effective field theories~\cite{RevModPhys.77.1423,Neubert:1993mb}, Hamiltonian approaches~\cite{PhysRevC.81.035205,Brodsky:1997de,Carbonell:1998rj,Keister:1991sb,PhysRevD.58.094030,PhysRevC.89.055205,PhysRevC.79.055203,Biernat2011}, and various kinds of effective phenomenological quark models, which are often based on quark-quark interactions similar to the Cornell potential~\cite{Eichten:1975bs}. 

In this paper we report on significant progress made in the description of dynamical quark mass generation in a framework called the ``Covariant Spectator Theory'' (CST)~\cite{Gross:1969eb,Gross:1982}. The CST is related to the BSDS formalism, from which it can be constructed, but it can also be viewed as a relativistic quark model, because its quark-quark interaction kernel contains a covariant phenomenological generalization of a linear confining potential, which is added to a one-gluon-exchange (OGE) kernel. 

As the name ``CST'' indicates, one of its most important features is that it is relativistically covariant, and this strict covariance must be preserved in all modifications of the kernel or in approximations to the full CST equations (also called the ``Gross equations''). These equations take the form of integral equations and are formulated in Minkowski space, as a consequence of which numerous singularities are encountered in the integrations over intermediate loop momenta. The zero-components of these loop momenta are then integrated by calculating only the residues of quark propagator poles, because the residues of the other poles (those in the kernel) tend to cancel. The remaining loop integrals are three-dimensional (but still covariant), which means that CST can also be categorized as a quasi-potential theory.

We have already shown that the CST approach is very capable of providing an excellent description of the masses of heavy and heavy-light mesons, even for highly excited states. However, our more ambitious aim is to make this framework self-consistent by including also the self-interaction of the quarks through the same kernel that describes the interaction between different quarks. This is equivalent to determining the dressed quark propagator, which can be expressed in terms of a dynamic quark mass (the ``mass function'') and a wave function renormalization, both dependent on the quark four-momentum.  

The calculation of the quark self-energy makes it feasible to implement an important constraint of QCD, namely the dynamical breaking of chiral symmetry. This is a key ingredient for a realistic description of the properties of light mesons, in particular the pion. How this can be done in CST, at least in principle, has been shown by Gross and Milana~\cite{Gross:1991te,Gross:1991pk,GMilana:1994} already in the early 1990s, and later by Gross and {\c{S}avkl\i}~\cite{Savkli:2001os}. We have investigated this issue further more recently~\cite{Biernat:2014jt,pionff:2014,pionff:2015,Biernat:2014xaa,Biernat:2012ig}, and found important constraints on the Lorentz structure of the kernel. 

However, this previous work always assumed a simplified kernel in which the full one-gluon exchange was approximated by a constant, which is clearly not sophisticated enough to account for the meson spectrum of Refs. \cite{Leitao:2017mlx,Leitao:2017it,Leitao2017,Leitao:2014}. It is the aim of this work to remove this simplification and to calculate the quark mass function and wave function renormalization from a kernel that includes the OGE mechanism exactly. If reasonable solutions can be obtained, it may indicate that the final goal of a completely self-consistent quark model in CST is within reach. To our knowledge no other quasi-potential approach has achieved such a degree of consistency. 

This paper is organized as follows: in Sec.~\ref{sec:I} we introduce the general definitions for the dressed quark propagator, as well as its relation to the self-energy and to the interaction kernel through the Dyson equation in its CST manifestation. 

Section~\ref{sec:III} analyzes the quark self-energy for the OGE kernel, and first discusses problems arising from overlapping poles in the quark and gluon propagators, which seem to render basic CST assumptions invalid, and presents a solution to these problems. It then shows in detail the calculation of the self-energy for the gluon-exchange kernel in a general linear covariant gauge, with special attention to the dependence of the results on the gauge parameter. We find that the gap equation, which determines the constituent quark mass, is completely independent of the gauge. However, off-shell quantities like the mass function away from the on-shell point, do depend on the gauge.
On the other hand, it turns out that the solution of the problems mentioned before leads to a preference of one particular gauge in CST. 

In Sec.~\ref{sec:ckernel} the calculation of the self-energy due to the constant kernel is presented, and we establish a useful relation between the constant kernel and the OGE self-energy. 
The overall result from both the constant and OGE kernels is discussed in Sec.~\ref{sec:CplusOGE}. We find good agreement between our mass function from the complete kernel and results from lattice QCD calculations in the spacelike region. For timelike quark momenta the gauge-dependence is more pronounced, except near the on-shell point. We summarize our findings and present our conclusions in Sec.~\ref{sec:VI}.

\section{Quark self-energy and the dressed quark propagator}  \label{sec:I}

{\subsection{General definitions}

 The  dressed quark propagator $S(p)$ for a bare (current) quark mass $m_0$ and four-momentum $p$, is given by the non-linear equation
\bea
S(p)&=&S_0(p){- } S_0 (p) Z_2 \Sigma({\slashed p}) S(p)\,,
\label{eq:DE}\eea
where  { $S_0(p)=\left(m_0-\slashed{p}- \mathrm i\epsilon\right)^{-1}$} is the { bare} quark propagator, $Z_2$ is a renormalization constant (to be defined shortly), and $\Sigma(\slashed{p})$ is the quark self-energy given by
\bea
\Sigma (\slashed p)= -\mathrm i\int\frac{\mathrm d^4k}{(2\pi)^4} {\cal V}(p,k) S(k)\,,\label{eq:DSselfenergy}
\eea
with ${\cal V}(p,k)$ the interaction kernel. By iterating Eq.~(\ref{eq:DE}) the dressed quark propagator $S(p)$ can be written as an infinite series
\bea
S(p)&=&S_0(p) \sum_{n=0}^\infty \left[-Z_2 \Sigma({\slashed p}) S_0(p)\right]^n
\nonumber\\
&=&\frac1{m_0-\slashed{p}+Z_2\Sigma(\slashed{p})- \mathrm i\epsilon}\,. \label{eq:quarkprop}
\eea
Because $\slashed{p}^2=p^2$, $\Sigma({\slashed p})$ has the form
\bea
\Sigma(\slashed{p})=A(p^2)+\slashed{p}\,B(p^2)\, , \label{eq:se}
\eea
and the dressed propagator (\ref{eq:quarkprop}) can be written
\bea
S(p)=\frac{Z(p^2)}{M(p^2)-\slashed{p}-\mathrm i\epsilon}=
\frac{Z (p^2) [M(p^2)+ \slashed{p}]}{M^2(p^2)-p^2-\mathrm i\epsilon} \, .\qquad
\label{eq:prop}
\eea
The wave function renormalization $Z(p^2)$ and the mass function $M(p^2)$ are given by
\bea 
Z(p^2)&=&\frac{1}{1-Z_2B(p^2)}\label{eq:Z} \, ,\\
M(p^2)&=&Z(p^2) \left[ m_0+Z_2A(p^2) \right] \, , \label{eq:M}
\eea
respectively.
The quark will be on-shell with a dressed (constituent) mass $m$ if the gap equation
\bea
M(m^2)=m\label{eq:oscondition}
\eea
is satisfied, {\it i.e.}, if $S({p})$ has a real pole at $p^2=m^2$. In terms of the on-shell quantities $A_0\equiv A(m^2)$ (and similarly for $B$ and $Z$), the on-shell condition~(\ref{eq:oscondition}) becomes
\bea
m=m_0 +Z_2(A_0+m B_0)\, ,  
\eea
conveniently written
\bea
\frac{m-m_0}{m}=Z_2\left(\frac{A_0}{m}+B_0\right)\, . \label{eq:masseq}
\eea
The existence of a real mass pole of the dressed quark propagator is one of the central assumptions of the CST approach. It should be stressed that the fact that a single quark may be on mass-shell does not contradict quark confinement in this approach. In CST, quark confinement is realized through the special properties of our confining kernel, which never allows  both quark and antiquark (in meson states) or all three  quarks (in baryon states) to be on-shell at the same time~\cite{Savkli:2001os}. 

\subsection{Expansion near the on-shell point and CST self-energy}\label{sec:ZR}
Near the quark pole, after carefully expanding the quantities around $p^2=m^2$ in order to obtain the correct residue, the dressed quark propagator (\ref{eq:prop}) becomes
\bea
S(p)\simeq\frac{Z_0(m+\slashed{p})}{(1-2mM'_0)(m^2-p^2-\mathrm i\epsilon)} = \frac{Z_2(m+\slashed{p})}{m^2-p^2-\mathrm i\epsilon} \, ,\nonumber\\
\label{eq:8a}
\eea
where $Z_2$, the renormalization constant generally introduced in Eq. ~(\ref{eq:DE}), is here in CST given at the on-shell quark mass point $p^2=m^2$ and includes the residue of the quark pole (see also, for example, Chapter 11 of Ref.~\cite{Gross:1993zj}). This renormalization constant depends on parameters used to regularize divergent integrals.   In our approach these integrals are regularized by form factors, and hence $Z_2$ depends on the form factor parameters.  In addition, our $Z_2$ is defined at the quark mass pole $m$, so we will have a different $Z_2$ for each flavor of quark. The implications of this observation are discussed briefly at the end of Sec.~\ref{sec:seinFeynman} below, but a study of implications of the flavor ({\it i.e.} $m$) dependence of $Z_2$ is deferred to future work.

From Eqs.~(\ref{eq:Z}) and (\ref{eq:M}) we obtain
\bea
M'_0&=&\frac{\mathrm d M(p^2)}{\mathrm d p^2}\bigg|_{p^2=m^2}
=Z_0 Z_2(A'_0+mB'_0) \, ,
\eea
and hence an expression for $Z_2$,
\bea
Z_2&\equiv&\frac{Z_0}{1-2mM'_0}
\nonumber\\
&=&\frac1{1-Z_2 B_0-2m Z_2 (A'_0+mB'_0)}\, .\qquad \label{eq:ZRformula}
\eea

In the CST, the self-energy $\Sigma (\slashed p)$ is given by 
\bea
\Sigma (\slashed p)= -\mathrm i\int_{k0} {\cal V}(p,k) S(k)\,,\label{eq:DEk0}
\eea
where, $\mathcal V (p,k)\equiv \mathcal V (p,k;P)$ is the interaction 
kernel, which we take to be of the general form
\bea
\mathcal V (p,k)= \sum_{K}  V_{K}^{\mu\nu}(p,k)\Theta^K_\mu\otimes\Theta^{K}_\nu\left[\frac14\sum_a\lambda_a\otimes \lambda_a\right]
\, . \nonumber\\\label{eq:kernelstrcuture}
\eea
 Here, $V_K^{\mu\nu}(p,k)$ ($K=s,p,v$) is the momentum-dependent part, the $\Theta^K$'s ($\Theta^s=\mathbf {1},\Theta^p=\gamma^5$, \mbox{$\Theta^{v}_{\mu}=\gamma_\mu$}) are Lorentz-covariant operators acting in Dirac space, and the $\lambda_a$'s are the usual Gell-Mann matrices of $\mathrm {SU}(3)_\text{color}$. The action of the interaction kernel (\ref{eq:kernelstrcuture}) as operator is defined as\footnote{In colorless states (or for the self-energy) the color factor $\frac14\sum_a\lambda_a\lambda_a$ reduces to $\frac43$.}
 \bea
 {\cal V}(p,k) S(k)\equiv\frac43\sum_{K} V_K^{\mu\nu}(p,k)\Theta^{K}_{\mu} S(k)\Theta^{K}_{\nu} \,.\label{eq:Vdecomp1}
 \eea
The explicit form of the kernel will be specified shortly. The notation \lq \lq $k0$'' in (\ref{eq:DEk0}) indicates the CST prescription for performing the $k_0$  integration in the complex $k_0$ plane. It amounts to averaging the residues of the quark propagator poles in the upper and lower $k_0$ half-plane, and neglecting all residues of poles in the kernel (for more details see Ref.~\cite{Biernat:2014jt}). Therefore
\bea
 \mathrm i\int_{k0} \equiv  \mathrm i\int\frac{\mathrm d^4k}{(2\pi)^4}\,\bigg|_{\footnotesize\begin{array}{l}k_0\;\text{propagator} 
\cr \text{poles only}\end{array}}=-\frac12\sum_{\footnotesize\begin{array}{c}\text{propagator}
\cr \text{pole terms}\end{array}}\int_{\bf k} \, ,\nonumber\\\label{eq:pp}
\eea
where
\bea
\int_{\bf k}& \equiv & \int  \frac{\mathrm d^3 {\bf k}}{(2\pi)^3} \frac{m}{E_k}\,. 
\eea 
Then, Eq.~(\ref{eq:DEk0}) becomes
\bea
\Sigma (\slashed p)= \frac {Z_2}{ 2}\sum_{\sigma=\pm}\int_{\bf k} \mathcal V (p,{ \hat k_\sigma })  \Lambda (\hat k_\sigma ) \,,\label{eq:DEk01}
\eea
with
\bea
{ \Lambda (\hat k_\sigma)=\frac{m+\hat {\slashed{k}}_\sigma}{2m},}
\eea
and $\sigma=\pm$ labels the positive- and negative-energy on-shell momenta (corresponding to the positions of the quark propagator poles),  $\hat k_\sigma=(\sigma E_k,{\bf k})$, with \mbox{$E_k=\sqrt{m^2+{\bf k}^2}$}. 

Notice that, similar as the self-energy (\ref{eq:DSselfenergy}), the CST self-energy (\ref{eq:DEk01}) effectively includes all iterations of convoluted and aligned rainbow diagrams. The difference between (\ref{eq:DSselfenergy}) and (\ref{eq:DEk01}) is that the latter involves the self-energy $\Sigma(\slashed k)$ under the integral \emph{only} at the on-shell values $k=\hat k_\pm$, whereas (\ref{eq:DSselfenergy}) involves it at \emph{all} $k$ values. Thus, the two self-consistent equations for the CST self-energy are the mass gap and the $Z_2$ equations, Eqs.~(\ref{eq:oscondition})  and~(\ref{eq:ZRformula}), respectively. They fix the on-shell values $A_0$ and $B_0$ of the $A$ and $B$ functions, while Eq.~(\ref{eq:DEk01}) determines their off-shell behavior at arbitrary $p^2$.

The factor $Z_2$ in front of the integral in Eq.~(\ref{eq:DEk01}) appears for the following reason: because the Feynman diagram for the self-energy has all external lines ``amputated'', we only obtain a factor $Z_2=\sqrt{Z_2} \sqrt{Z_2}$ from the internal quark (a factor $\sqrt{Z_2}$ is associated with each quark line either entering or leaving an interaction vertex). This will lead to the renormalization of the couplings in the interaction kernel. Because a consistent renormalization procedure requires that all amplitudes be redefined by multiplying by factors of  $\sqrt{Z_2}$ for each external line, the additional factors are included in Eq.~(\ref{eq:DE}), and thus in the series expansion (\ref{eq:quarkprop}),\footnote{The first term in the series (\ref{eq:quarkprop}), with $n=0$, corresponds to the case when there is no interaction vertex and hence no factor of $Z_2$.} yielding an overall renormalization factor $Z_2^2$. 

\subsection{Interaction kernel}\label{sec:kernel}
We choose the interaction kernel (\ref{eq:kernelstrcuture}) as 
 \bea
 {\cal V}(p,k) &=&{\cal V}_{\ell}(p,k)+{\cal V}_{\mathrm g}(p,k)+{\cal V}_{\mathrm c}(p,k)\nonumber\\&=& 
 \left\{\left[\left(1-\lambda\right)\left(\mathbf{1}\otimes \mathbf{1}+\gamma^5\otimes\gamma^5\right)-\lambda \gamma^\mu\otimes\gamma_\mu \right]V_{\ell} (p,k)\right.\nonumber\\& &-\left.\gamma_\mu\otimes\gamma_\nu \left[\Delta_{\mathrm g}^{\mu\nu}(q^2)V_{\mathrm g}(p,k)+\Delta_{\mathrm c}^{\mu\nu}(q^2)V_{\mathrm c}(p,k)\right] \right\}\nonumber\\&&\times\left[\frac14\sum_a\lambda_a\otimes \lambda_a\right]\,,\label{eq:Vdecomp}
 \eea
 where ${\cal V}_{\ell}$ is a covariant generalization of a linear
confining potential, ${\cal V}_{\mathrm g}$ is the short-range effective OGE  interaction, and ${\cal V}_{\mathrm c}$ a covariant form of a constant potential, to be specified below. The parameter $\lambda$ ($0\leq\lambda\leq1$) in the linear confining part parametrizes the mixing between scalar-pseudoscalar and vector Lorentz structures. The $\Delta^{\mu\nu}$'s are covariant factors that depend on the gauge parameter $\xi$. In Ref.~\cite{Biernat:2014xaa} we prove that this kernel satisfies the axial-vector Ward-Takahashi identity and is therefore consistent with chiral symmetry and its spontaneous breaking.

In the present work we set $\lambda=0$, {\it i.e.}, we use a pure scalar-pseudoscalar linear confining part. In Ref.~\cite{Leitao:2017mlx} we found that the kernel~(\ref{eq:Vdecomp}) with $\lambda\approx0$ gives a very good description of the heavy and heavy-light meson spectrum. Setting $\lambda=0$ in this work has the advantage that the linear confining part gives no contribution to the CST self-energy (\ref{eq:DEk01}). This is because  this part does not contribute to the $A$-function of~(\ref{eq:se}), and in the $B$-function the contributions from the scalar and pseudoscalar parts of the linear confining kernel cancel~\cite{Biernat:2014xaa}. This simplifies the problem substantially while maintaining consistency with the meson  calculations, because we only have to calculate the contributions from the OGE and constant part of the kernel to the self-energy. The general case $\lambda\neq0$ will be considered in future work.
}

\section{Self-energy from one-gluon-exchange kernel}
\label{sec:III}
\subsection{Complications with self-energy calculations in CST}
\label{sec:prescriptionC}
Before carrying out any specific calculations, we focus on a central problem with extending the CST to the calculation of  self-energies.  

\begin{figure}[h]
 \centering
\includegraphics[width=\linewidth]{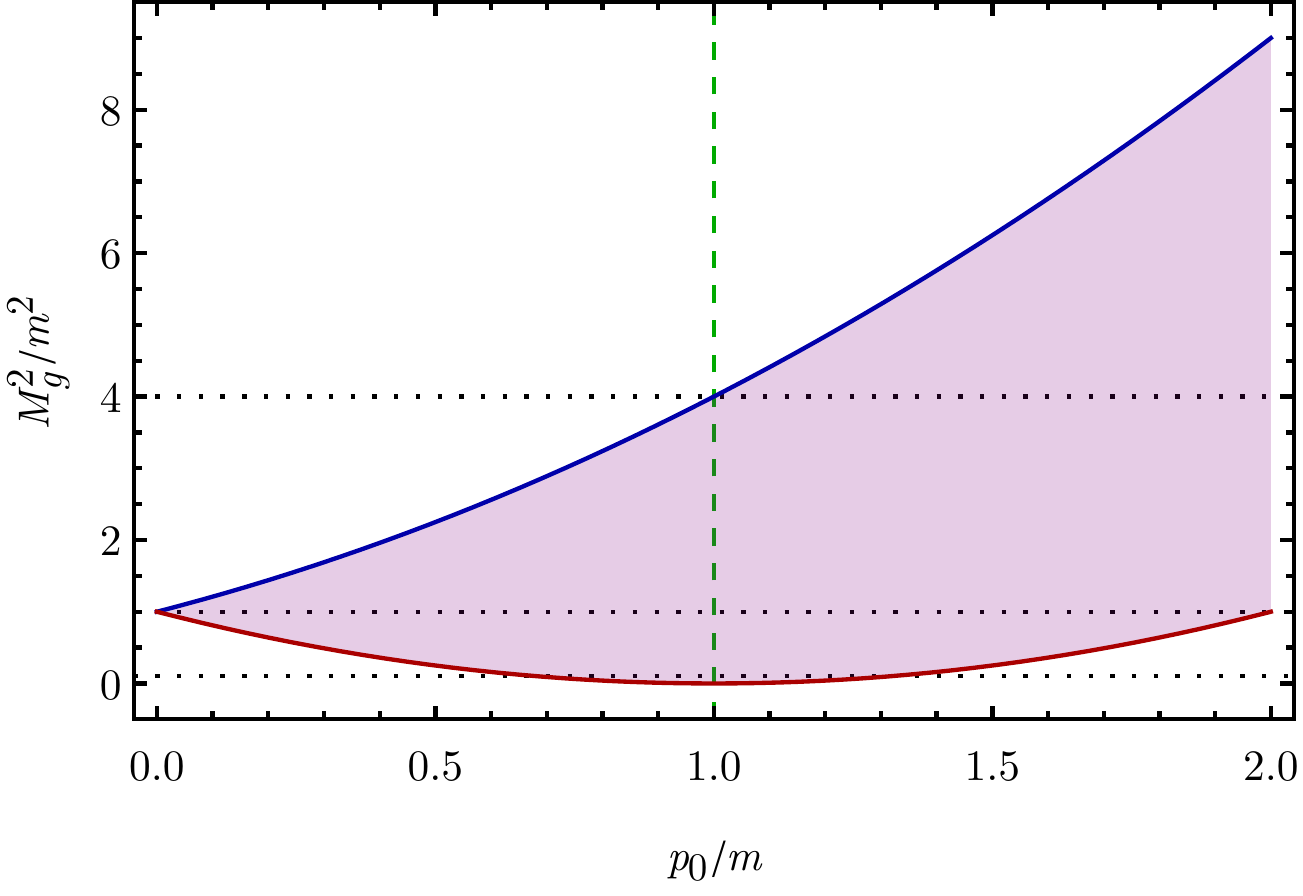}
 \caption{Singularity-free region (shaded in purple) of $M_{\rm g}^2/m^2$ vs. $p_0/m$. The region is limited by the curves $(1+\frac{p_0}{m})^2$ (blue solid line) and $(1-\frac{p_0}{m})^2$ (red solid line). The black dotted reference lines are for $M_{\rm g}^2/m^2 = 4, 1, 0.1$.  The green dashed vertical line marks the point $p_0=m$.}
 \label{fig:q2lines}
\end{figure}

Consider the self-energy for a quark of mass $m$ and a gluon of mass $M_{\rm g}$. 
Ignoring spin, it is of the form
 \bea
 \Sigma(p^2)&\propto& -\mathrm i\int\frac{\mathrm d^4k}{(2\pi)^4}\frac1{(D_{\mathrm q}-\mathrm i\epsilon)({ D}_{\mathrm g}-\mathrm i\epsilon)}  \, . \label{eq:A1}
 \eea
In the quark's rest frame, where $p=p^{\rm r}\equiv \{p_0,{\bf 0}\}$ with $p_0>0$ (it should be noted that in this paper, we use a rather loose notation where $p_0\equiv p_0^{\rm r}$ always refers to the zeroth component of $p^{\rm r}$, and not of the general $p$), the denominators are
\bea
D_{\rm q}-\mathrm i\epsilon&=&(E_k-k_0-\mathrm i\epsilon)(E_k+k_0-\mathrm i\epsilon) \, ,
\nonumber\\
{ D}_{\rm g}-\mathrm i\epsilon&=&M_{\rm g}^2-q^2-\mathrm i\epsilon
\nonumber\\
&=&({\cal E}_k+p_0-k_0-\mathrm i\epsilon)({\cal E}_k-p_0+k_0-\mathrm i\epsilon)\, ,\qquad \label{eq:DmD0}
\eea
 and 
\bea
{\cal E}_k=\sqrt{M_{\rm g}^2+{\bf k}^2}\ .
\eea
To calculate the residues of the quark poles at $k_0=\sigma E_k$,
 $q^2$ has to be replaced by
 \bea
q_\sigma^2=m^2+p_0^2- 2\sigma p_0 E_k\, . \label{eq:qsigma2}
\eea 
Because $E_k\ge m$, $D_{\rm g}$ has zeros (and hence the gluon propagator has  singularities) when
\begin{equation}
M_{\rm g}^2 \ge  (m+p_0)^2 \quad \mbox{or}  \quad
M_{\rm g}^2 \le (m-p_0)^2\, .
\end{equation}
On the other hand, $D_{\rm g}$ does not hit any singularities if
\bea
(m-p_0)^2 \le M_{\rm g}^2 \le (m+p_0)^2\, .
\eea
This singularity-free region is illustrated in Fig.~\ref{fig:q2lines}.  Note that $D_{\rm g}$ is not singular at the point $p_0^2=m^2$ as long as $0\le M_{\rm g}^2\le 4 m^2$.

The presence of a singular region for $M_{\rm g}^2\le(m-p_0)^2$ is an undesirable feature of the CST calculation of self-energies.  If $k_0=E_k$ ({\it i.e.}, at the positive-energy quark pole) this singularity can be traced  to the zero of the gluon denominator at the point $E_k-{\cal E}_k= p_0$.  This zero is the condition for the overlap of the gluon and quark poles in the lower half plane. 
Similarly, if $k_0=-E_k$ (at the negative-energy quark pole), the condition for the overlap of the gluon and quark poles in the upper half plane occurs at ${\cal E}_k-E_k= p_0$. In either case, when quark and gluon poles overlap it is no longer justified to keep only one and neglect the other, and therefore the basic idea of CST is not applicable. It could appear unavoidable that the full BSDS calculation has to be performed. 
     
However, CST has so many positive features, such as three-dimensional integrations (as compared to four-dimensional integrations in BSDS), that it is worthwhile looking for a less drastic alternative. The most important point in this context is that a CST calculation of the quark self-energy, {\it i.e.}, one in which only the quark propagator poles are kept, makes it still possible to satisfy the constraints of chiral symmetry on the two-quark sector in a simple and elegant way, while maintaining complete consistency between the one- and two-quark problems. If the self-energy included the contributions of other poles than the ones of the quark propagator, this mechanism would no longer work. Instead of paying the high price of the complexity of four-dimensional integrations, we prefer to modify the kernel, in such a way, that overlapping poles are avoided both in the one- and the two-body CST equations and consistency with chiral symmetry is maintained at the same time.

Note that the discussion above assumes that the gluon propagator is dressed in the simplest possible way: generating a fixed momentum independent mass $M_{\rm g}$. Our discussion would be altered if the dressed gluon propagator had only cuts (and no poles) along the real axis \cite{Alkofer:2003jj,Strauss:2012dg}, or if the gluon mass function were momentum dependent with its own gap equation \cite{Aguilar:2015bud}.  Study of these possibilities is well beyond the scope of this work, and these and many issues involving the dressed gluon propagator can be addressed when the gluon mass function is studied using the CST. Until then, in this first work, we assume the dressed gluon has a fixed mass and discuss how the shortcomings of the CST can be addressed under this assumption.

To find a sensible and practical solution to the problem as outlined above, we draw on past experience. The presence of unwanted singularities has been encountered before in the application of the CST to two- and three-nucleon scattering \cite{Gross:1991pm,Gross:2008ps}.  Here, when the same nucleon is on-shell before and after the exchange of a meson, the meson momentum transfer $q^2\le0$ and no singularities are encountered.  But when alternate nucleons are on-shell, the momentum transfer $q^2$ of the exchanged meson propagators can become positive, and singularities appear.  One of these singularities is a production singularity, expected and needed when the energies are large enough to produce physical mesons in the intermediate state, but the other is a spurious singularity that arises from negative energy nucleons, just as those that appear in the self-energy when $M_{\rm g}^2\le (m-p_0)^2$.   In the two-body problem it can be shown that these singularities are canceled by higher order diagrams (in that case the crossed boxes), so they are unphysical and need not be carefully evaluated.  

Three prescriptions for the treatment of the meson propagators have been applied to in the $NN$ problem.  The first (A) need not be considered here. The second (B) is to leave $q^2$ unaltered and carefully integrate over the singularities (this was done in Ref.~\cite{Gross:1991pm}), and the third (C) (used in Ref.~\cite{Gross:2008ps}) is to replace $q^2$ in the denominator of the propagator by $-|q^2|$. These are the two choices we face here, and we have found, as will be shown shortly, that Prescription C most faithfully reproduces the physics expected from the contributions of OGE to the quark propagators.  

We emphasize that this prescription does not alter any calculation for which  $q^2<0$, which is always the case for the one-channel CST equations used for the study of the heavy and heavy-light meson spectrum.  It does, however, have a significant effect on the calculation of the self-energy.  One of the byproducts of this paper is a demonstration that Prescription C gives good results for self-energies, and hence is preferred to Prescription B. (For an in-depth discussion of these issues, see Appendix~B of Ref.~\cite{Gross:2008ps}.)

There is a second problem that arises in the CST self-energy calculation: a form factor inserted to provide convergence for the $k$ integrals for the quark self-energy, will not provide such convergence if it is purely a function of $q^2$.  The problem occurs when $p_0^2=0$.  At that point, $q^2$ becomes a constant (equal to $m^2$) independent of $k$, and thus a function solely of $q^2$ cannot regularize the integral. We have considered this issue at some length, and the only way that we have discovered to avoid infinities in $A_{\rm g}$ and $B_{\rm g}$ at $p_0\to 0$  is to use a form factor that depends on the covariant variable $(p\cdot k)^2/(k^2p^2)$ instead of on $q^2$.  This regularization form factor will be introduced below.

 \subsection{Effective one-gluon-exchange kernel in general linear covariant gauge}

 The effective OGE kernel in arbitrary gauge is given by
 \begin{eqnarray}
 \mathcal V_{\rm g}(p,k)&=&4\pi {\left[\frac14\sum_a\lambda_a\otimes \lambda_a\right]}\alpha_{\rm s} \, g(y)\, \gamma_\mu\otimes\gamma_\nu\;\frac{\Delta^{\mu\nu}_{\mathrm g}(q^2)}{(-q^2)}
  \nonumber\\
 \label{eq:OGEgauge0}
\end{eqnarray}
where 
 \begin{eqnarray}
 \Delta_{\mathrm g}^{\mu\nu}(q^2)= Q_{\rm g}(q^2) \left(\mathrm g^{\mu\nu}-\frac{q^\mu q^\nu}{q^2}\right)+\xi L_{\rm g}(q^2)\frac{q^\mu q^\nu}{q^2}\,,
 \label{eq:gaugefactor}\nonumber\\
\end{eqnarray}
is the gauge factor with $Q_{\rm g}(q^2)$ and $L_{\rm g}(q^2)$ transversal and longitudinal gluon dressing functions, respectively, $\xi$ is the gauge parameter, $\alpha_{\rm s}$ is the \emph{unrenormalized} strong coupling constant, and $g(y)$ is a regularization form factor depending on the covariant variable $y^2$ 
\bea
y^2=\frac{(p\cdot k)^2}{p^2k^2} \to \frac{E_k^2}{m^2} \, ,
\eea
where the limiting form emerges in the rest frame of $p$ with $k$ on-shell, so that $p=p^{\rm r}$ and $k^2=m^2$.  The form factor will be normalized to 1 for on-shell $k$ at the point ${\bf p}={\bf k}=0$.  

For the gluon dressing functions we use 
\begin{eqnarray}
Q_{\rm g}(q^2)=  L_{\rm g}(q^2)= -\frac{q^2}{M_{\rm g}^2+|q^2|},\label{eq:ansaetzedressing}
\end{eqnarray}
such that
\begin{eqnarray}
 \frac{\Delta_{\mathrm g}^{\mu\nu}(q^2)}{(-q^2)}=  \frac{1}{M_{\rm g}^2+|q^2|}\left[\mathrm g^{\mu\nu}-(1-\xi)\frac{q^\mu q^\nu}{q^2}\right] , \,
\end{eqnarray}
and the kernel becomes
\begin{eqnarray}
 \mathcal V_{\rm g}(p,k)&=&4\pi{\left[\frac14\sum_a\lambda_a\otimes \lambda_a\right] \alpha_{\rm s}} \, g(y)\, \gamma_\mu\otimes\gamma_\nu\nonumber\\&&\times\frac{1}{M_{\rm g}^2+|q^2|}\left[\mathrm g^{\mu\nu}-(1-\xi)\frac{q^\mu q^\nu}{q^2}\right]  . \,
  \label{eq:OGEgauge}
\end{eqnarray}

The choice (\ref{eq:ansaetzedressing}) effectively implements the Prescription C  discussed in Sec.~\ref{sec:prescriptionC}, by both giving the gluon a finite mass $M_{\rm g}$ and replacing $q^2\to -|q^2|$, which removes the singularity in the gluon propagator. 

Because  $L(q^2)\neq1$ the longitudinal part gets also dressed, and one might object that this violates the Slavnov-Taylor identity for the gluon self-energy. This is indeed true for the OGE part of (\ref{eq:OGEgauge}) \emph{alone}. We should stress, however, that our OGE is only part of a phenomenological kernel that also includes a constant and a linear confining part, which are not necessarily purely transverse and they might also include longitudinal dressing effects. The gauge dependence of the confining part, which can be viewed as an effective multi-gluon exchange, is, however, far from clear, and computing the gluon self-energy self-consistently from the interaction kernel would go beyond the scope of this work.

\subsubsection{Self-energy in Feynman-'t Hooft gauge} \label{sec:seinFeynman}

First we use only the ${\rm g}^{\mu\nu}$ term of the kernel (\ref{eq:OGEgauge}), which corresponds to choosing $\xi=1$, {\it i.e.}, the Feynman-'t Hooft gauge. Working in the timelike region ($p^2>0$) in a frame where $p =p^{\rm r}$, the quark self-energy is 
\begin{eqnarray}
Z_2\Sigma_{\rm g}(\slashed p)\equiv Z_2\Sigma_{\rm g}^{\xi=1}(\slashed p)&=& 
\frac{8\pi }{3}Z_2^2\alpha_{\rm s}\sum_\sigma \int_{\bf k} \frac{g(y)(4m-2\slashed {\hat k }_\sigma)}{M_{\rm g}^2+|q_\sigma^2|}  \nonumber\\
&=& Z_2^2 \alpha_{\rm s} [\bar{A}_{\rm g}(p_0^2)+\slashed{p}\bar{B}_{\rm g}(p_0^2)]\, ,\label{eq:sigma1}
\end{eqnarray}
where we used the fact that $y^2$ is independent of the sign of $E_k$. The \emph{reduced} scalar self-energy functions  become
\bea
\bar{A}_{\rm g}(p_0^2)&=& \frac{32\pi }{3} \,m\sum_\sigma\int_{\bf k}
\frac{g(y)}{D_\sigma}
\, ,
\nonumber\\
\bar{B}_{\rm g}(p_0^2)&=&- \frac{16\pi }{3} \sum_\sigma\int_{\bf k} \frac{\sigma E_k}{p_0}\frac{g(y)}{D_\sigma}\, ,
 \label{eq:Feynman}
\end{eqnarray}
where the denominators are
\bea
D_\sigma&=&M_{\rm g}^2+|q^2_\sigma|\,. 
\eea
Note that here, and in this following discussion, we have removed a factor of $Z_2^2\alpha_{\rm s}$ from the structure functions defined in Eq.~(\ref{eq:se}); to avoid confusion these reduced structure function are written with a bar ({\it i.e.} $ Z_2 A \equiv Z_2^2 \alpha_{\rm s}\bar{A} $ and  $ Z_2 B \equiv Z_2^2 \alpha_{\rm s}\bar{B}$).  The overall factor of $Z_2^2 \alpha_{\rm s}$ in (\ref{eq:sigma1}) is the \emph{renormalized} strong coupling constant
\bea
\alpha_{\rm s}^{\rm r}\equiv Z_2^2 \alpha_{\rm s}\, . \label{eq:alphas}
\eea
The renormalization (\ref{eq:alphas}) of $\alpha_{\rm s}$ arises from a factor of $\sqrt{Z_2}$ attached to each quark line either entering or leaving an interaction vertex.  Because the invariants $A$ and $B$ were defined using the expansion (\ref{eq:se}), they include only \emph{one} of the factors of $Z_2$, with the other factor $Z_2$ multiplying the LHS of Eq.~(\ref{eq:sigma1}) coming from the definition~(\ref{eq:quarkprop}).

Notice that $\alpha_{\rm s}^{\rm r}$ depends on the constituent quark mass $m$ through $Z_2$, but this might not be the only dependence on the quark mass. We know that $\alpha_{\rm s}^{\rm r}$ runs with $q^2$, and in a calculation of this kind the average value of $q^2$ depends on the quark mass. Hence a calculation that includes this running should predict how the average value of $\alpha_{\rm s}^{\rm r}$ depends on $m$, and the $m$ dependence of $Z_2$ could be incorporated into this dependence. In this work we only consider the chiral limit, and $\alpha_{\rm s}^{\rm r}$ here is given at the quark mass in the chiral limit. Once we have obtained results for heavy quark flavors -- planned for a subsequent paper we will be able to predict how $\alpha_{\rm s}^{\rm r}$ varies with the quark mass. This will provide an indirect way to assess the physical content of the model.

 \subsubsection{Self-energy in general linear covariant gauge}

Next we generalize the self-energy calculation to general linear covariant gauges. For values $\xi \neq1$, also the $q^\mu q^\nu$ term of the kernel (\ref{eq:OGEgauge}) contributes to the self-energy. This adds the term $(1-\xi)Z_2 \Delta \Sigma_{\rm g}(\slashed{p})$ to the Feynman-'t~Hooft-gauge contribution $Z_2\Sigma_{\rm g}(\slashed{p})$, such that the self-energy in arbitrary gauge reads 
\bea
Z_2 \Sigma_{\rm g}^{\xi}(\slashed{p})=Z_2 \Sigma_{\rm g}(\slashed{p})+ (1-\xi)Z_2 \Delta \Sigma_{\rm g}(\slashed{p})\, , \label{eq:Sigmaxi}
\eea
with 
\bea
Z_2 \Delta{\Sigma}_{\rm g}(\slashed{p})&=&-\frac{8\pi}{3} \alpha_{\rm s}^{\rm r}\sum_\sigma\int_{\bf k}\frac{g(y)}{D_\sigma}
\bigg\{\frac{\slashed{q}_\sigma (m+\slashed{\hat k}_\sigma) \slashed{q}_\sigma}{q_\sigma^2} \bigg\}\, .
\nonumber\\ \label{eq:deltasigmag}
\eea
 Using $q_\sigma = \hat k_\sigma-p$ in the frame where $p =p^{\rm r}$,
the term in braces in (\ref{eq:deltasigmag}) reduces to
\bea
\frac{\slashed{q}_\sigma (m+\slashed{\hat k}_\sigma) \slashed{q}_\sigma}{q_\sigma^2} =m+\gamma^0 \left(\sigma E_k +\frac{2{\bf k}^2p_0}{q^2_\sigma}\right)\, .
\eea
Combining this with the results from the Feynman-'t~Hooft gauge gives 
\bea
\bar{A}_{\rm g}^\xi(p_0^2)&=&\frac14[3+\xi] \bar{A}_{\rm g}(p_0^2) \, ,
\nonumber\\
\bar{B}_{\rm g}^\xi(p_0^2)&=&\frac12[3-\xi] \bar{B}_{\rm g}(p_0^2) -[1-\xi] \bar{R}_{\rm g}(p_0^2)\, , \label{eq:AelBel}
\eea
where the superscript $\xi$ indicates that $A$ and $B$ are calculated in an arbitrary gauge, and the remainder term is given by
\bea
\bar{R}_{\rm g}(p_0^2)=\frac{16\pi}{3} \sum_\sigma\int_{\bf k}\frac{g(y){\bf k}^2}{D_\sigma q_\sigma^2}\, .\qquad\quad \label{eq:Rfun}
\eea

\subsection{Analysis of the integrals}   
\label{sec:startnew}

We turn now to the evaluation of the integrals (\ref{eq:Feynman}) and (\ref{eq:Rfun}). Because the integrands involve the absolute value of $q^2$, their careful analysis is somewhat intricate. Here we only present the results, and refer to Appendix~\ref{app:startnew} for details. 

In (\ref{eq:Feynman}) and (\ref{eq:Rfun}), it is convenient to change the radial integration variable from $|{\bf k}|$  to $y=E_k/m$, and to introduce the dimensionless variable $s=p_0^2/m^2$, such that the dressed quark mass $m$ can be scaled out:  
\bea
\frac{\bar{A}_{\rm g}(s\,m^2)}{m} &=& \int_1^\infty \mathrm dy\sqrt{y^2-1}\,g(y) {\cal A}_{\rm g}(s,y)\,,\nonumber\\
\bar{B}_{\rm g}(s\,m^2)
&=&\int_1^\infty \mathrm dy\sqrt{y^2-1} \, g(y) {\cal B}_{\rm g}(s,y)\,,
\nonumber\\
\bar{R}_{\rm g}(s\,m^2)
&=&  \int_1^\infty \mathrm dy\sqrt{y^2-1}\,g(y){\cal R}_{\rm g}(s,y)\,.
\label{eq:32}
\eea
For $s>0$, {\it i.e.}, for timelike momenta $p$, the interval of the $y$ integration is divided into two regions by the point $y_0\equiv\frac{1+s}{2\sqrt s}$: 

 In Region 1, where $y\leq y_0$, the  functions in the integrands of (\ref{eq:32}) are 
\bea
{\cal A}^1_{\rm g}(s,y)&=&
\frac{16\chi_1}{3\pi(\chi_1^2-4y^2s)}\,,
\nonumber\\
{\cal B}^1_{\rm g}(s,y)&=&
-\frac{16y^2}{3\pi(\chi_1^2-4y^2s)}\,,
\nonumber\\
{\cal R}_{\rm g}^1(s,y)&=&
\frac{8(y^2-1)[(1+s)\chi_1+4y^2s]}{3\pi\rho_+^2\rho_-^2(\chi_1^2-4y^2s)} \,, \qquad
\label{eq:region1}
\eea
where
$\rho_\sigma^2=q_\sigma^2/m^2$ and $\chi_1\equiv m_{\rm g}^2+1+s $ with $m_{\rm g}=M_{\rm g}/m$. In Region 2 the functions in the integrands read

\bea
{\cal A}^2_{\rm g}(s,y)&=&
\frac{16\chi_2}{3\pi(\chi_2^2-(1+s)^2)} \, ,
\nonumber\\
{\cal B}^2_{\rm g}(s,y)&=&
-\frac{8y(1+s)}{3\pi \sqrt{s}(\chi_2^2-(1+s)^2)} \, ,
\nonumber\\
{\cal R}_{\rm g}^2(s,y)&=&
\frac{8(y^2-1)(1+s)(\chi_2+2y\,\sqrt{s})}{3\pi \rho_+^2\rho_-^2(\chi_2^2-(1+s)^2)}\, ,\qquad
\label{eq:region2}
\eea
where $
\chi_2\equiv m_{\rm g}^2+2y\,\sqrt{s}$. Notice that the asymptotic points $s=0_+$ and $s=\infty$ lie in Region 1, while the on-shell point $s=1$ lies in Region 2.

When $p$ is spacelike ($s<0$) there is only one region, which we call Region 0, and the functions in the respective integrands are
\bea
{\cal A}^0_{\rm g}(s,y)&=&
\frac{16}{3\pi(m_{\rm g}^2+\rho_{\rm c}^2)}\, ,
\nonumber\\
{\cal B}^0_{\rm g}(s,y)&=&
0\, ,
\nonumber\\
{\cal R}_{\rm g}^0(s,y)&=&
\frac{8(y^2-1)(1+s)}{3\pi\rho_{ c}^4(m_{\rm g}^2+\rho_{\rm c}^2)}\, .  \label{eq:ABR0}
\eea
where 
$\rho_{c}^2=\sqrt{(1+s)^2-4y^2s}$.

\begin{figure}
 \centering
 \includegraphics[width=\linewidth]{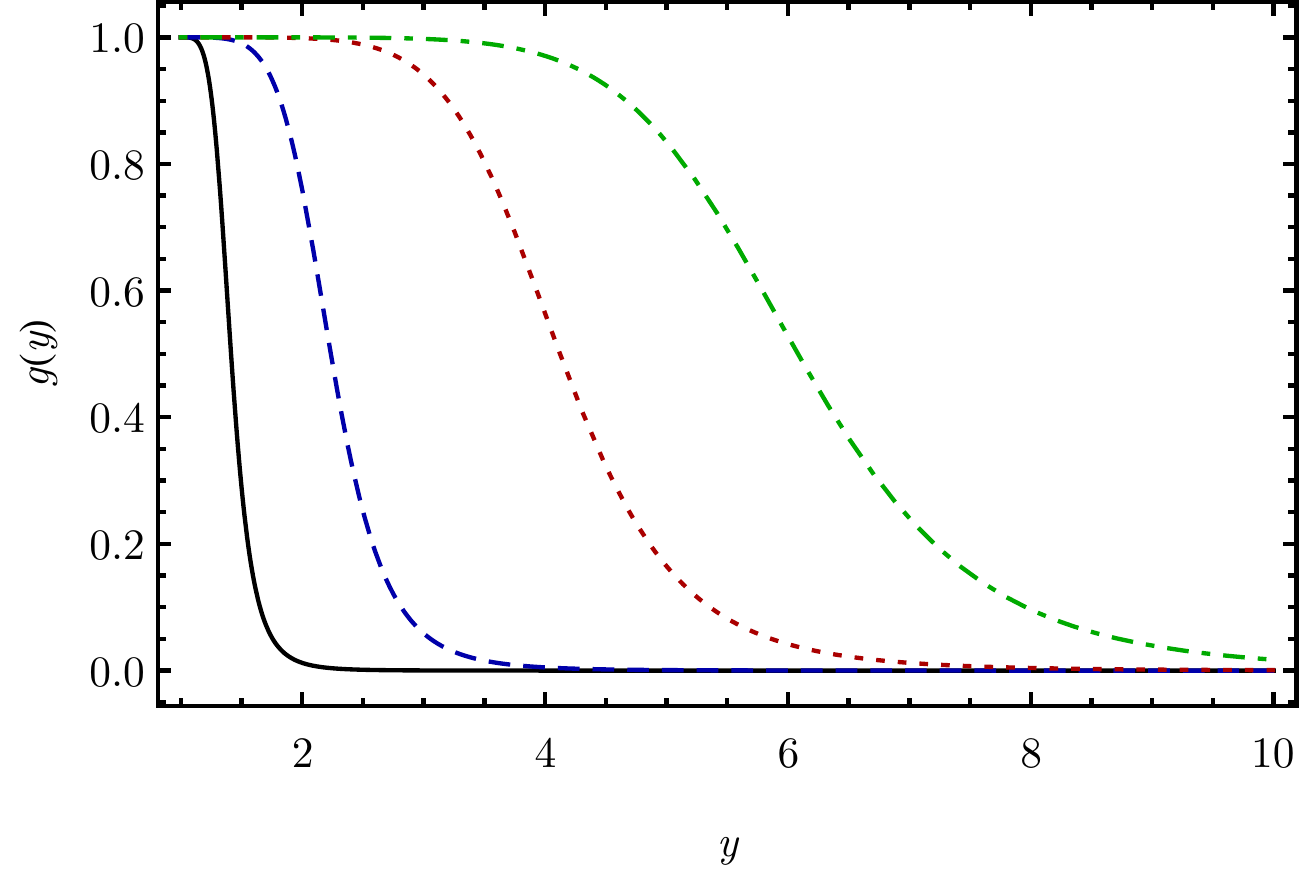}
 \caption{The form factor $g(y)$ vs. the dimensionless variable $y$ for $\lambda_{\rm g}=1$ (solid black), $\lambda_{\rm g}=2$ (dashed blue),  $\lambda_{\rm g}=4$  (dotted red), and  $\lambda_{\rm g}=6$  (dot-dashed green).  In all cases, $n=4$.   }
 \label{fig:hg}
\end{figure}

Comparing Regions 0 and 1, we observe that ${\cal B}_{\rm g}$ is not continuous at $s=0$. At first, this seems to be a 
serious problem, because a discontinuous self-energy can hardly be considered acceptable. 
However, a simple solution is to choose the gauge parameter $\xi = 3$ (known in the literature as the Yennie gauge~\cite{Fried:1958zz}), such that $\bar{B}_{\rm g}^{\xi}$ of Eq.~(\ref{eq:AelBel}) remains continuous at $s=0$ despite the discontinuity of ${\cal B}_{\rm g}$.\footnote{Another possibility to make $\bar{B}_{\rm g}^{\xi}(p_0^2)$ continuous would be to impose constraints on the
form factor $g(y^2)$. However, this possibility is not further pursued here.} 
It is worth emphasizing that this issue about discontinuity is not an unescapable feature of CST itself, but only a consequence of 
Prescription C for dealing with kernel singularities. 

The large-$s$ behavior of the self-energy invariants is independent of the gluon mass, as expected. The asymptotic behavior of the integrals
helps to find reasonable values for the parameters of the form factor $g$. We chose the form
\bea
g(y)&=&
\frac{\lambda_{\rm g}^{2n}}{\lambda_{\rm g}^{2n}+(y^2-1)^n} \, ,
 \label{eq:hg}
\eea
for which the asymptotic integrals converge only when $n\ge3$.  We use $n=4$ in our numerical computations, but the results are quite insensitive to the precise value of $n$.  The behavior of $g$ for different values of $\lambda_{\rm g}$ is shown in Fig.~\ref{fig:hg}.

\subsection{Gauge independence at $s=1$}

 \begin{figure}[t]
 \centering
 \includegraphics[width=\linewidth]{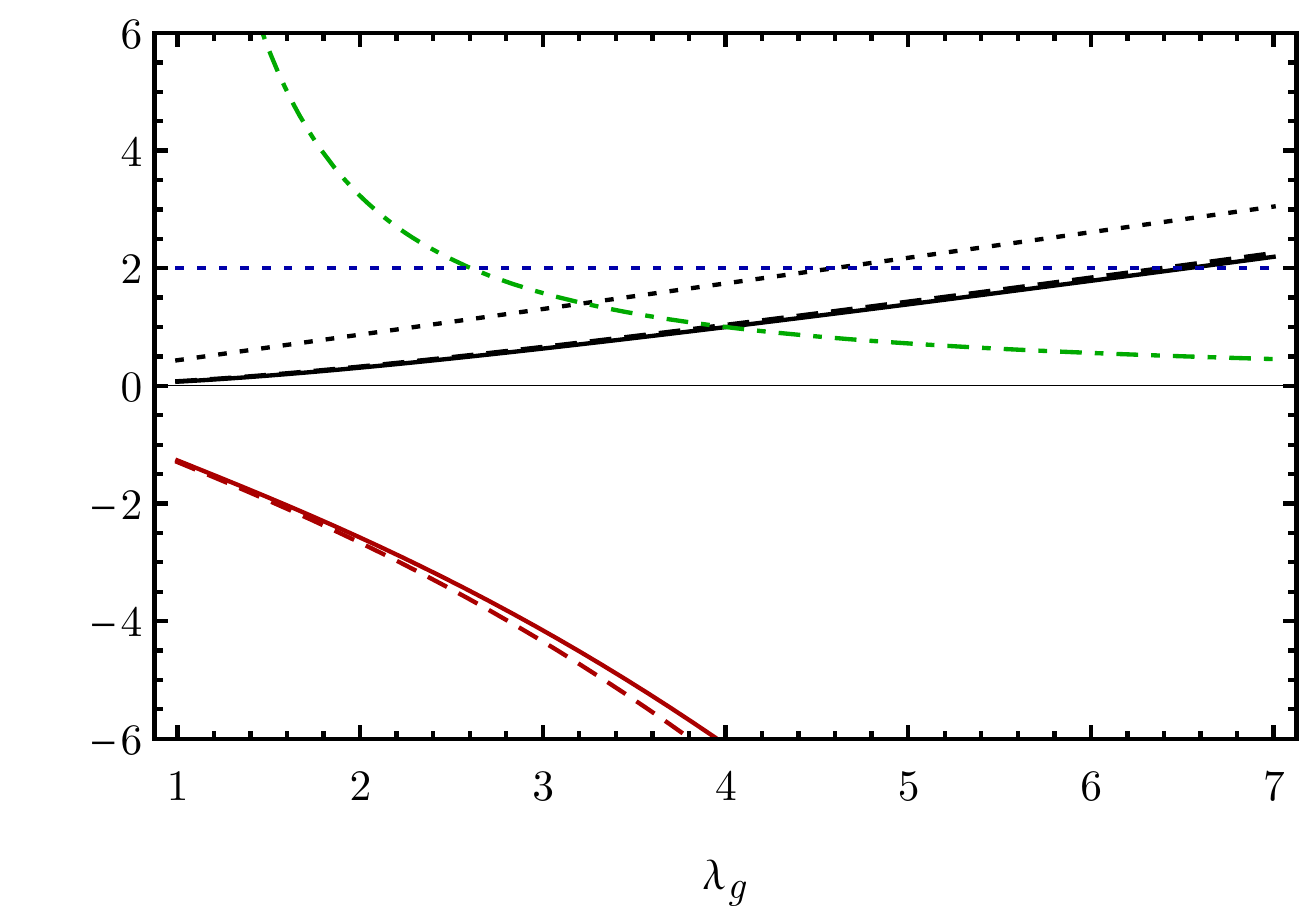}
 \caption{The dimensionless structure function $T_{\rm g}$ (black linear lines with values greater than zero) and $T_{\rm g}^{\rm B}$ (red lines with values less than zero) vs. $\lambda_{\rm g}$ for $n=4,m_{\rm g}^2=4$ (solid lines) and $n=3, m_{\rm g}^2=4$ (dashed lines).  The black dotted line slightly larger than $T_{\rm g}$ ($n=4, m_{\rm g}^2=4$) is $T_{\rm g}$ ($n=4; m_{\rm g}^2=0$). The dot-dashed green line is $\alpha_{\rm s}^{\rm r}$ computed from $T_{\rm g}$ ($n=4, m_{\rm g}^2=4$).  All of these curves are independent of the gauge. For comparison, the blue dotted line shows the value of 2.  }
 \label{fig:TGvsLg}
\end{figure}

Next we look at the on-shell point $s=1$ ($p^2=m^2$), for which the functions in the integrands of (\ref{eq:32}) lie entirely in Region 2. 
The gap equation (\ref{eq:masseq}) for the OGE kernel can be written 
\bea
\frac{m-m_0}{m}=\alpha_{\rm s}^{\rm r}T_{\rm g}^\xi \, , \label{eq:massgapgeneralTg}
\eea
in terms of the function 
\bea
T_{\rm g}^\xi&\equiv&\frac{\bar{A}^\xi_{\rm g}(m^2)}{m}+\bar{B}^\xi_{\rm g}(m^2)
\nonumber\\&=& 
\int_1^\infty {\mathrm d}y\sqrt{y^2-1}\,g(y) {\cal T}_{\rm g}^\xi(y)\, ,
\label{eq:Tgxi}
\eea
where the factor in the integrand is
\bea
{\cal T}_{\rm g}^\xi(y)&\equiv&\frac14[3+\xi]{\cal A}^1_{\rm g}(1,y)+ \frac12[3-\xi]{\cal B}^1_{\rm g}(1,y)\nonumber\\&&-[1-\xi] {\cal R}_{\rm g}^1(1,y)
\nonumber\\&=& 
\frac{16(m_{\rm g}^2+y)}{3\pi[(m_{\rm g}^2+2y)^2-4]}
\nonumber\\&=&
{\cal A}_{\rm g}^1(1,y)+{\cal B}^1_{\rm g}(1,y)={\cal T}_{\rm g}(y)\,.\quad 
\label{eq:AgplusBg}
\eea
This factor is independent of the gauge, and, because $T_{\rm g}^\xi=T_{\rm g}$ immediately follows from (\ref{eq:Tgxi}), so is the gap equation. 

We note that when ${\cal T}_{\rm g}^\xi(y)$ is calculated with Prescription B instead of C, using the results from Eq.~(\ref{eq:regionx}), we obtain
\bea
 {\cal T}^{{\rm B}\xi}_{\rm g}(y)&\equiv&\frac14[3+\xi]{\cal A}^{\rm B}_{\rm g}(1,y)+ \frac12[3-\xi]{\cal B}^{\rm B}_{\rm g}(1,y)\nonumber\\&&-[1-\xi] {\cal R}_{\rm g}^{\rm B}(1,y)
\nonumber\\&=&\frac{16(m_{\rm g}^2-2+y^2)}{3\pi[(m_{\rm g}^2-2)^2-4y^2]}
\nonumber\\&=&
{\cal A}^{\rm B}_{\rm g}(1,y)+{\cal B}^{\rm B}_{\rm g}(1,y)={\cal T}_{\rm g}^{\rm B}(y)\,,\quad 
\label{eq:AgplusBgx}
\eea
which is {also} gauge-parameter independent. 
 
This gauge independence at the on-shell point is a general feature of the CST. To see this we multiply the self-energy of Eq.~(\ref{eq:Sigmaxi}) by the on-shell projection operator $\Lambda({\hat p}_\sigma)$, and, using $\Lambda({\hat p}_\sigma)\slashed{\hat p}_\sigma 
=\Lambda({\hat p}_\sigma) m$, we obtain
\bea
\frac{Z_2}{\alpha_{\rm s}^{\rm r}}\Lambda({\hat p}_\sigma)\Sigma_{\rm g}^\xi(\slashed{\hat p}_\sigma)&=&
\left(\frac{m+\slashed{\hat p}_\sigma}{2m}\right)\Big[\bar{A}^{\xi}_{\rm g}(m^2)+\slashed {\hat p}_\sigma\bar{B}^{\xi}_{\rm g}(m^2)\Big]\nonumber\\&=&\Lambda( {\hat p}_\sigma)m \, T_{\rm g}^\xi=\Lambda({\hat p}_\sigma)m \, T_{\rm g}\, ,
 \label{eq:proja&B}
\eea
which is independent of $\xi$ because of Eqs.~(\ref{eq:AgplusBg}) or~(\ref{eq:AgplusBgx}), irrespective of which  prescription is applied to handle the kernel singularities. In Appendix~\ref{app:gaugeindep} we show that this gauge independence is a consequence of the fact that the $q^\mu q^\nu$-term of the kernel drops out of the CST Dyson equation~(\ref{eq:DE}) at the on-shell point. Only unprojected or off-shell results are sensitive to the gauge, and this limits the impact the choice of gauge can have on any calculation.
 
To study the conditions under which we obtain solutions of the gap equation, it is convenient to examine the dependence of $T_\mathrm{g}$ on details of our model, such as the parameters $\lambda_{\rm g}$ and $n$ of the form factor $g(y)$, and the method to handle singularities of the kernel. 
Figure~\ref{fig:TGvsLg} displays  $T_{\rm g}$ (using Prescription C) and $T_{\rm g}^{\rm B}$ (using Prescription B) for different values of the gluon mass $m_\mathrm{g}$, of the form factor parameter $\lambda_{\rm g}$ and for different exponents $n$.  When $T_{\rm g} \simeq 2$, the mass gap equation~(\ref{eq:massgapgeneralTg}) in the chiral limit ($m_0=0$) is solved by a renormalized strong coupling constant
\bea
\alpha_{\rm s}^{\rm r} \simeq 0.5 \quad\mbox{for}\quad\lambda_{\rm g}\simeq 7\,, \label{eq:alpha1} 
\eea
which corresponds to a typical OGE strength of calculations of the meson spectrum.\footnote{Since we carried the color factor $\frac43$ through the calculation, $\frac 43 \alpha_{\rm s}^{\rm r}$ should be compared with the value of the strong coupling constant from the meson-spectrum paper~\cite{Leitao:2017mlx}.} 

From the figure we can draw four important conclusions: (i) the solution to the mass gap equation is sensitive to the range parameter $\lambda_{\rm g}$, and for $\lambda_{\rm g}\simeq 7$ Prescription C gives a satisfactory solution if $m_{\rm g}\simeq2$; (ii) the solution is insensitive to $n$; (iii) the solution depends on $m_{\rm g}$, but is qualitatively unchanged even for the extreme case $m_{\rm g}=0$; and (iv) we do not find a solution to the mass gap equation if Prescription B is used.

\subsection{Quark mass and wave functions from one-gluon-exchange kernel for $s<0$} \label{Sec:Massoge}
Now we turn to a discussion of the quark mass function and wave function renormalization from the OGE kernel at negative $s$. We treat $\lambda_{\rm g}$ as an adjustable parameter, but the other parameters are held fixed at
 \bea 
n&=&4 \, ,
\nonumber\\
m_{\rm g}&=&2\, ,
\nonumber\\
m&=&0.3\;{\rm GeV}\, ,
\nonumber\\
m_0&=&0\,, 
  \label{eq:hgparas}
\eea
and Prescription C is used throughout. 

\begin{figure}
 \center
 \includegraphics[width=\linewidth]{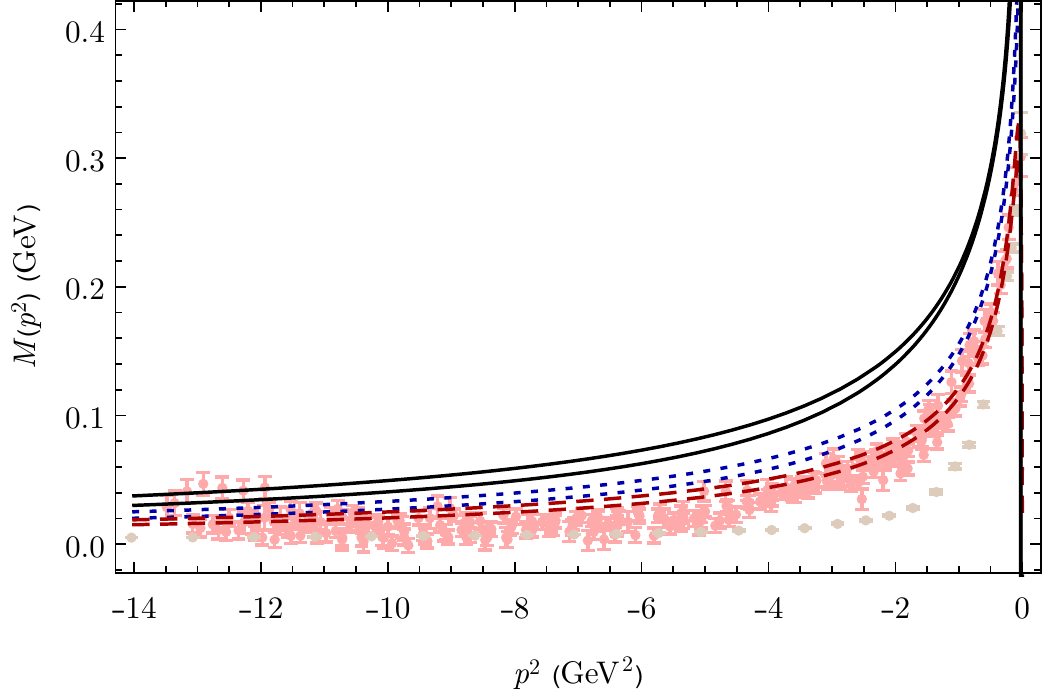}
 \caption{ The mass function $M$ (in GeV) vs. $p^2$ (in GeV$^2$) for $\xi = 3$ (two solid black lines), $\xi=1$ (two dotted blue lines), and $\xi = 0$ (two dashed red lines).   In every pair, the larger result is for $\lambda_{\rm g}=5$ and the smaller for $\lambda_{\rm g}=3$.  The other parameters are given in Eq.~(\ref{eq:hgparas}). The lattice QCD data are taken from Refs. \cite{Bowman:2005vx} (red data points) and \cite{Oliveira:2018lln} (brown data points).   }
 \label{fig:MassOGE}
\end{figure}

The quark mass function 
\bea
M(p^2)= \alpha_{\rm s}^{\rm r} \bar{A}_{\rm g}^\xi(p^2) Z(p^2)
\eea
with
\bea
Z(p^2)=\frac{1}{1-\alpha_{\rm s}^{\rm r}\bar{B}_{\rm g}^\xi(p^2)}\,\label{eq:ZOGE}
\eea 
for $s<0$ is sensitive to the gauge and to the parameter $\lambda_{\rm g}$. However, Fig.~\ref{fig:MassOGE} shows that the dependence of our mass function on $\lambda_{\rm g}$ is quite weak, and that our results in Landau gauge ($\xi=0$) agree remarkably well with the  lattice data of Ref.~\cite{Bowman:2005vx}.

There are caveats one should keep in mind when comparing our mass and renormalization functions with lattice QCD data. The latter are only available for spacelike quark momenta in Landau gauge, so rigorously we should also only compare our Landau-gauge results with these data. On the other hand, for spacelike momenta our results do not depend much on the gauge, so a comparison with our results obtained in other gauges, in particular the Yennie gauge, makes sense as long as one stays away from the region close to $p^2=0$. However, one can also argue that the mass function and the wave function renormalization are not observables, and that therefore agreement or disagreement with the lattice data would not decide whether our CST results are reasonable or not.  A real test requires the use of the dressed CST quark propagator in the calculation of genuine observables, such as in the calculation of meson properties. This is planned for the near future.

The renormalized strong coupling constant $\alpha_{\rm s}^{\rm r}$ for the parameters used in Fig.~\ref{fig:MassOGE} are
\bea
\alpha_{\rm s}^{\rm r} &=& 0.722 \quad\mbox{for}\quad\lambda_{\rm g}=5\, ,
\nonumber\\
\alpha_{\rm s}^{\rm r} &=& 1.577\quad\mbox{for}\quad\lambda_{\rm g}=3\,.\label{eq:alphasrvalues}
\eea
 Note that in these examples the values of $\lambda_{\rm g}$ are smaller and the values of $\alpha_{\rm s}^{\rm r}$ are larger than the ones in Eq.~(\ref{eq:alpha1}).

\begin{figure}[tbh]
 \center
\includegraphics[width=\linewidth]{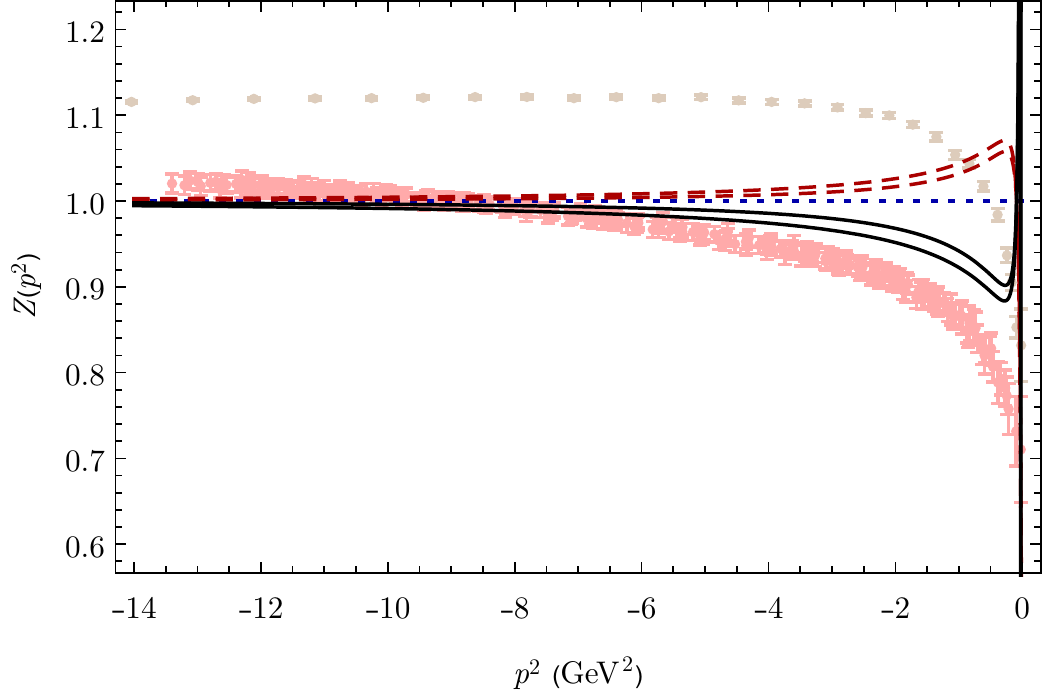}
 \caption{ The wave function renormalization $Z$ vs. $p^2$ (in GeV$^2$) for $\xi = 3$ (two solid black lines: upper with $\lambda_{\rm g} = 5$; lower $\lambda_{\rm g}=3$), $\xi=1$ (dotted blue line; $Z=1$ for all $\lambda_{\rm g}$), and $\xi = 0$ (two dashed red lines: lower with $\lambda_{\rm g} = 5$; upper $\lambda_{\rm g}=3$). The other parameters are given in Eq.~(\ref{eq:hgparas}). The lattice QCD data are taken from Refs. \cite{Bowman:2005vx} (red data points) and \cite{Oliveira:2018lln} (brown data points). }
 \label{fig:ZOGE}
\end{figure}

Figure~\ref{fig:ZOGE} shows the wave function renormalization $Z$, Eq.~(\ref{eq:ZOGE}), for the cases shown in Fig.~\ref{fig:MassOGE}.  Only the Yennie gauge gives  a shape for $Z$ that dips below 1 for $p^2\lesssim-8$ GeV$^2$, as predicted by the lattice data. However, in all gauges a  zero in $\bar{R}_{\rm g}(p^2)$ appears at $p^2=-m^2$ ($s=-1$) [recall  (\ref{eq:ABR0})], which differs from the behavior of the lattice data.  This might be corrected by using more realistic gluon dressing functions $Q_{\rm g}$ and $L_{\rm g}$ that include, for instance, a running gluon mass (see the discussion in the final section).

\section{Self-energy from a constant kernel}
 \label{sec:ckernel}
 When the CST self-energy is calculated from the OGE kernel only, the value of $\alpha_{\rm s}^{\rm r}$ turns out to be unnaturally large as compared to the approximate value known from experiment [see the discussion in Sec.~\ref{Sec:Massoge} and for the values see Eq.~(\ref{eq:alphasrvalues})]. As will be shown below, the presence of an additional constant in the kernel solves this issue and leads to realistic values of our model parameters.
\subsection{Constant kernel in general linear covariant gauge}
 The covariant constant vector kernel we consider in this section is of the form 
{ \bea
{\cal V}_{\rm c}(p, \hat k_\sigma)&=&\frac{C E_k}{2m}(2\pi)^3\delta^3\Big({\bf k}-\frac{m}{\sqrt{p^2}}\, {\bf p}\Big) %
h(p^2)h(m^2)\nonumber\\&&\times
\gamma_\mu\otimes\gamma_\nu \, \Delta_{\mathrm c}^{\mu\nu}(q_\sigma^2) \, , \label{eq:Ckernel}
\eea where $C$ is the unrenormalized strength of the interaction and the normalization of the form factor $h$ (specified below) is 
\bea
h(m^2)=1\, . \label{eq:hcnorm}
\eea
A proof of covariance of this kernel can be found in Appendix~\ref{app:covaVc}.

If the constant kernel is regarded as a correction to, or a partial substitution for the OGE contribution, it is gauge dependent as well, and (\ref{eq:Ckernel}) is the corresponding expression in general linear covariant gauge. The gauge-dependent factor is
 \begin{eqnarray}
\Delta_{\mathrm c}^{\mu\nu}(q^2)= Q_{\mathrm c}(q^2) \left(\mathrm g_{\mu\nu}-\frac{q_\mu q_\nu}{q^2}\right)+\xi L_{\mathrm c}(q^2)\frac{q_\mu q_\nu}{q^2}\, ,
 \label{eq:gaugefactorconstant}\nonumber\\
\end{eqnarray}
and $Q_{\mathrm c}(q^2)$ and $L_{\mathrm c}(q^2)$ are the transverse and longitudinal dressing functions, respectively. In principle, $L_{\mathrm c}(q^2)$ is determined from the longitudinal part of the OGE kernel through the Slavnov-Taylor identity, but, as already discussed above, this would go beyond the scope of this work. Instead, in this work we choose, for simplicity, $Q_{\mathrm c}(q^2)=L_{\mathrm c}(q^2)=1$.

\subsection{Self-energy and D$\chi$SB}

\subsubsection{Feynman-'t Hooft gauge}
Using the kernel (\ref{eq:Ckernel}) in Feynman-'t Hooft gauge \mbox{($\xi=1$)}, the quark self-energy is 
\bea
Z_2\Sigma_{\rm c}(\slashed{p})&\equiv&Z_2\Sigma^{\xi=1}_{\rm c}(\slashed{p})\nonumber\\&=&-\mathrm i\, Z_2C h(p^2)\int_{k0} \frac{\mathrm d^4k}{(2\pi)^4} \frac{E_k}{2m}(2\pi)^3
\nonumber\\
&&\qquad\times \delta^3\Big({\bf k}-\frac{m}{\sqrt{p^2}}\, {\bf p}\Big)%
\gamma^\mu S(k)\gamma_\mu
\nonumber\\
&=&-\mathrm i\, Z_2^2 C h(p^2)\int_{k0} \frac{\mathrm dk_0}{(2\pi)} \frac{1}{2} 
\left(\frac{4m-2\gamma^0k_0}{m^2-k_0^2-\mathrm i\epsilon}\right) 
\nonumber\\
&=& \bar {C}^{\rm r}m\, \frac{h(p^2)}{2}\sum_\sigma\left(1-\frac12\sigma \gamma^0\right)  =\bar {C}^{\rm r}m\,h(p^2)
\, ,\nonumber\\
 \label{eq:Sigmac}
\eea
where, in the second line the covariant expression has been evaluated in the frame $p=p^{\rm r}$, fixing \mbox{${\bf k}={\bf p}={\bf 0}$}, and in the last line the average of the contributions from the poles at \mbox{$k_0=\sigma m=\pm m$} is computed.   As in Eq.~(\ref{eq:alphas}), we write the final answer in terms of the scaled renormalized strength of the constant interaction 
\bea
\bar {C}^{\rm r}m \equiv Z_2^2\,C\, .
\eea 
We conclude that the invariant functions generated by the constant interaction are 
\bea
Z_2A_{\rm c}(p^2)&=&\bar {C}^{\rm r} m\,h(p^2) \, ,
\nonumber\\
Z_2 B_{\rm c}(p^2)&=&0\, . \label{eq:AcBc}
\eea  
Hence,
\bea
Z(p^2)&=&1 \, ,
\nonumber\\
M(p^2) &=& m_0+Z_2 A_{\rm c}(p^2)= m_0+ \bar {C}^{\rm r}\, m\,h(p^2)\, . \label{eq:ZandM} 
\eea
In view of condition (\ref{eq:hcnorm}) and the mass gap equation (\ref{eq:oscondition}), spontaneous chiral symmetry breaking requires $\bar {C}^{\rm r}=1$, a result we obtained before~\cite{Biernat:2014jt}.

\subsubsection{General linear covariant gauge} \label{sec:Cingauge}
In arbitrary gauge ($\xi\neq1$), the $q^{\mu}q^{\nu}$ term of the kernel  (\ref{eq:Ckernel}) contributes and the  self-energy includes the additional term $(1-\xi)Z_2\Delta \Sigma_{\rm c}(\slashed{p})$, where 
\bea
Z_2\Delta \Sigma_{\rm c}(\slashed{p})&=&\mathrm i\, Z_2C h(p^2)\int_{k0} \frac{\mathrm d^4k}{(2\pi)^4} \frac{E_k}{2m}(2\pi)^3\nonumber\\
&&\qquad\times  \delta^3\Big({\bf k}-\frac{m}{\sqrt{p^2}}\, {\bf p}\Big) 
\frac{\slashed{q} S(k)\slashed{q}}{q^2}
\nonumber\\
&=&\mathrm i\,Z_2^2C \,h(p^2)\int_{k0} \frac{\mathrm d k_0}{(2\pi)} \frac{E_k}{2m}  
\left[\frac{\gamma^0(m+\gamma^0k_0)\gamma^0}{m^2-k_0^2-\mathrm i\epsilon}\right] 
\nonumber\\
&=& - \bar {C}^{\rm r}m\, \frac{h(p^2)}8\sum_\sigma\left(1+\sigma\gamma^0 \right)
=-\frac14  \bar {C}^{\rm r}m\, h(p^2) \, .\nonumber\\
\label{eq:deltaS}
\eea
Comparing with the result in Feynman-'t Hooft gauge, Eq.~(\ref{eq:Sigmac}), we see that, in arbitrary gauge, $A_{\rm c}$ is modified by a  factor
\bea
A_{\rm c}^\xi&=&\frac14[3+\xi]A_{\rm c}
\, .\label{eq:A_B_ell}
\eea
Inserting this into the mass gap equation (\ref{eq:oscondition}) gives
\bea
\frac{m-m_0}{m}
&=& \frac14\bar {C}^{\rm r}(3+\xi) \, , \label{eq:massgapgeneral}
\eea
showing that, if $\bar {C}^{\rm r}$ is to satisfy the mass gap equation in an arbitrary gauge, it must itself be gauge dependent, {\it i.e.}, $\bar {C}^{\rm r}\to \bar {C}_\xi^{\rm r}$.  For spontaneous chiral symmetry breaking,
\bea
 \bar {C}_\xi^{\rm r}=\frac{4}{3+\xi} \, . \label{eq:CR}
\eea  

\begin{figure}[tbh]
 \centering
\includegraphics[width=\linewidth]{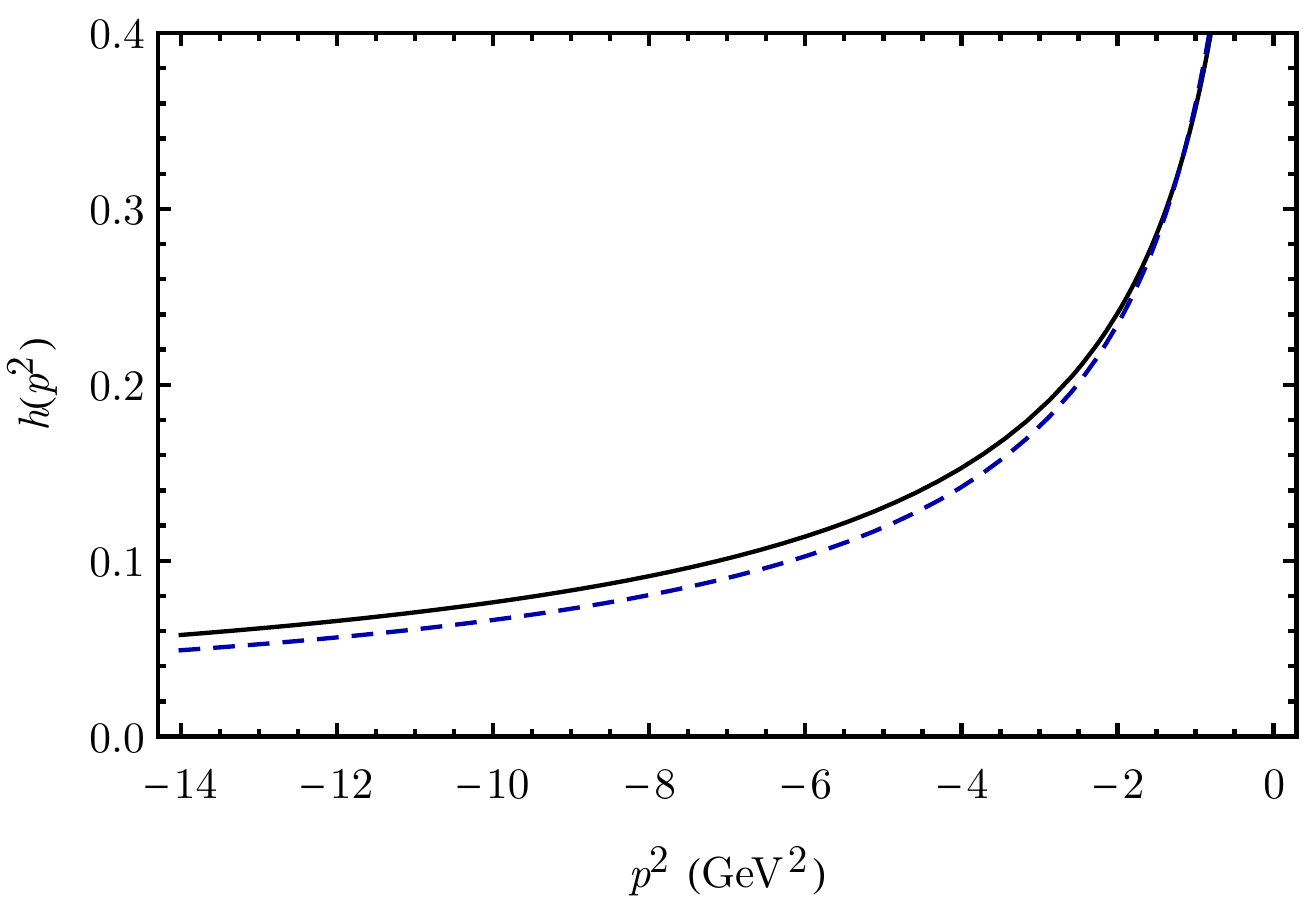}
 \caption{The form factor $h(p^2)$ plotted as a function of $p^2$ (in GeV$^2$).  The solid black line has $\lambda_{\rm g}=5$; the blue dashed line has  $\lambda_{\rm g}=3$.  The other parameters are given in Eq.~(\ref{eq:hgparas}).  }
 \label{fig:Aformfactor}
\end{figure}

\begin{figure}[tbh]
 \centering
\includegraphics[width=\linewidth]{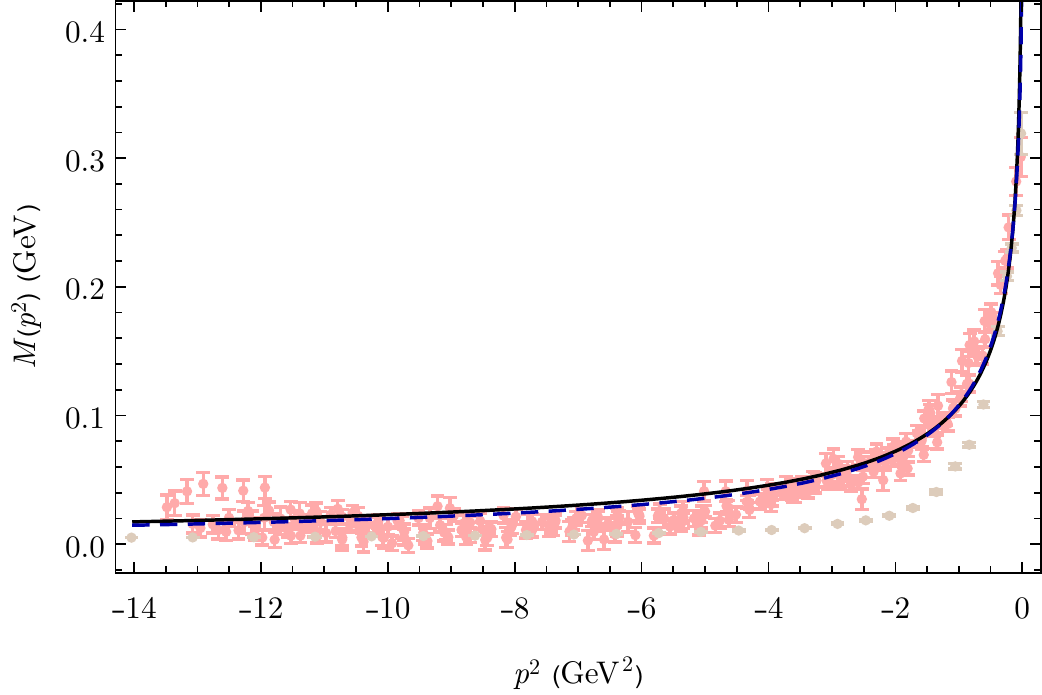}
 \caption{The mass function $M$ (in GeV) predicted by the form factor $h(p^2)$ plotted as a function of $p^2$ (in GeV$^2$).  The solid black line has $\lambda_{\rm g}=5$; the blue dashed line has  $\lambda_{\rm g}=3$.  The other parameters are given in Eq.~(\ref{eq:hgparas}). The lattice QCD data are taken from Refs. \cite{Bowman:2005vx} (red data points) and \cite{Oliveira:2018lln} (brown data points).   }
 \label{fig:MC}
\end{figure}

\subsection{Mass function from the constant kernel}

If the constant kernel is to supplement the OGE kernel, then it is appropriate to choose the form factor $h$ to be $\bar{A}_{\rm g}(p^2)$ normalized to unity at $p^2=m^2$, according to~(\ref{eq:hcnorm}):
\bea
h(p^2) = \frac{\bar{A}_{\rm g}(p^2)}{\bar{A}_{\rm g}(m^2)}\, .\label{eq:hc}
\eea
In this way the constant potential is entirely determined through the scalar part of the OGE self-energy. 
Using the parameters (\ref{eq:hgparas}) we consider the two cases
{\bea
\bar{A}_{\rm g}(m^2)&=& 0.26\,{\rm GeV}\qquad \lambda_{\rm g}=3\,,
\nonumber\\
\bar{A}_{\rm g}(m^2)&=& 0.603 \,{\rm GeV}\qquad \lambda_{\rm g}=5\, .
\eea}
The form factor for each of these cases in the region $s<0$  is shown in Fig.~\ref{fig:Aformfactor}.  One can see that it is almost independent of $\lambda_{\rm g}$. The mass function  for this form factor is shown in Fig.~\ref{fig:MC}.  Since the mass gap equation (\ref{eq:massgapgeneral}) holds in both cases, the prediction for the mass function, in any gauge, is obtained by multiplying the form factors in Fig.~\ref{fig:Aformfactor} by $m$,
\bea
M(p^2)=m\, h(p^2) \,. 
\eea
Our results look very similar to the lattice data of Ref.~\cite{Bowman:2005vx}.
 
 \section{Constant plus OGE self-energy} \label{sec:CplusOGE}

Next, we calculate the quark self-energy when the OGE and constant kernels are added together.  Since both $\alpha_{\rm s}^{\rm r}$ and $\bar {C}^{\rm r}_\xi$ were chosen to satisfy the mass gap equation separately, any linear combination of these contributions will also be a solution, suggesting that the combined result be written as 
\bea
Z_2 A(p^2)&=&{\frac14(3+\xi)\Bigg[\eta \,\alpha_{\rm s}^{\rm r}+(1-\eta)\frac{{ m} \bar {C}_\xi^{\rm r}}{\bar{A}_{\rm g}(m^2)} \Bigg]\bar{A}_{\rm g}(p^2)}
\nonumber\\
&=&\Bigg[\frac14(3+\xi)\eta \,\alpha_{\rm s}^{\rm r}+(1-\eta)\frac{{ m}}{\bar{A}_{\rm g}(m^2)} \Bigg]\bar{A}_{\rm g}(p^2)
\, ,
\nonumber\\
Z_2 B(p^2)&=&\eta\,\alpha_{\rm s}^{\rm r} \bar{B}^\xi_{\rm g}(p^2) \, , \label{eq:ABcg}
\eea
where $\eta$ is a mixing parameter and the constraint (\ref{eq:CR}) was applied in the expression for $Z_2 A(p^2)$. The results for the  quark wave function renormalization and the quark mass function are then obtained by inserting the expressions of (\ref{eq:ABcg}) into Eqs.~(\ref{eq:Z}) and~(\ref{eq:M}), respectively.

Since the strong coupling constant is roughly known from experimental data (we will assume  $\alpha_{\rm s}^{\rm p}= 0.5$ for the purposes of this paper), we choose $\eta$ to reproduce this value regardless of the choice of $\lambda_{\rm g}$.  Hence we define
\bea
\eta\equiv \eta(\lambda_{\rm g})= { \frac{\alpha_{\rm s}^{\rm p}}{\alpha_{\rm s}^{\rm r}(\lambda_{\rm g})}}\, .
\eea 
With this constraint on $\eta$, we can study the dependence of the mass function and of $Z$ on the scale parameter $\lambda_{\rm g}$, knowing that the experimental value $\alpha_{\rm s}^{\rm p}$ will always emerge. 

\begin{table}[tb]
\begin{minipage}{2.5 in}
\caption{Parameters for the constant plus OGE self-energy. In all cases $\alpha_{\rm s}^{\rm p}=0.5$.}
\begin{ruledtabular}
\begin{tabular}{l|ccc}
$\xi$ & 0 & 1 & 3 \\[0.05in]
$\lambda_{\rm g}$ & 3 & 2 &1.5  \\[0.05in]
$\eta(\lambda_{\rm g})\quad$ & { 0.317}  & { 0.155} & { 0.087} \\[0.05in] 
$(1-\eta)\bar {C}^{\rm r}_\xi$ & { 0.911} & { 0.845}& { 0.608} \\[0.05in]
\end{tabular}
\end{ruledtabular}
\label{tab:paras}
\end{minipage}
\end{table}

We conclude that the contribution from the constant kernel in Eq.~(\ref{eq:ABcg}) effectively decreases the strength $\alpha_{\rm s}^{\rm r}$ by a factor $\eta$, such that the  effective strength of the OGE contribution $\eta\alpha_{\rm s}^{\rm r}$ assumes the experimental value $\alpha_{\rm s}^{\rm p}$. Therefore, it seems as if the presence of a constant in the kernel somewhat corrects for the omission of the gluon-pole contributions in the CST self-energy calculation.         

\begin{figure}[h]
 \center
\includegraphics[width=\linewidth]{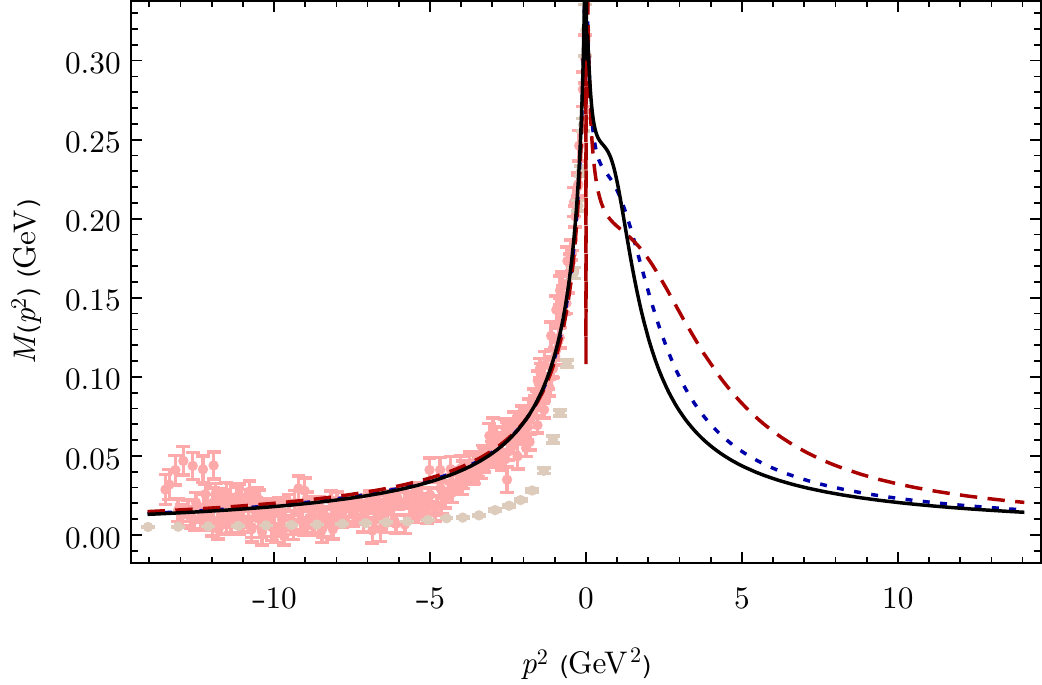}
\includegraphics[width=\linewidth]{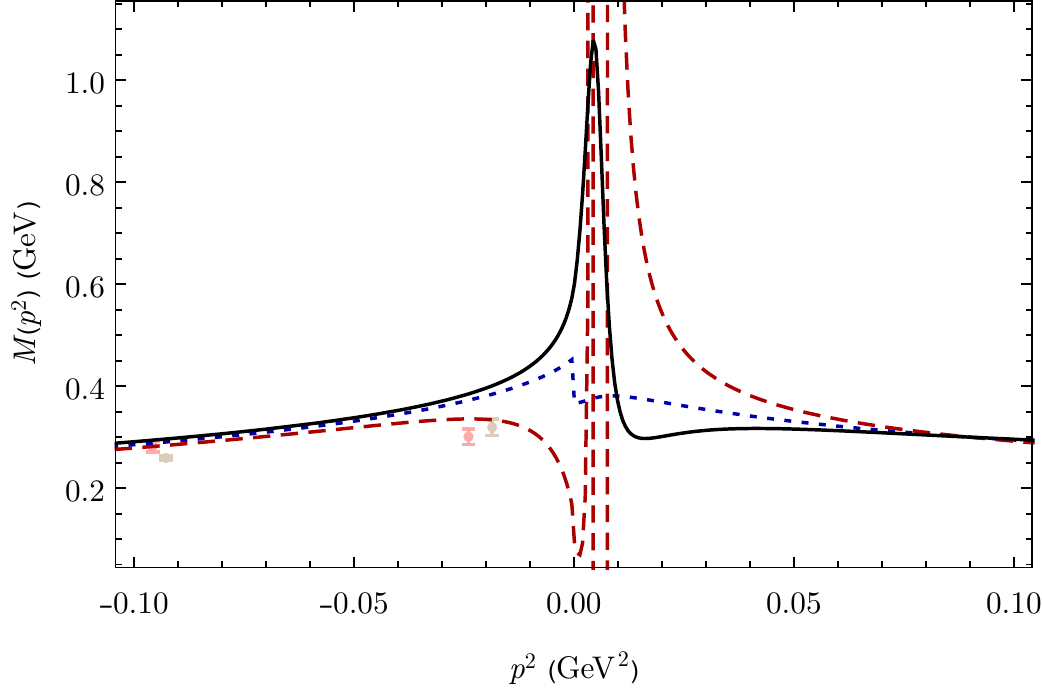}
 \caption{The mass function for the sum of constant and OGE kernels (in GeV) vs. $p^2$ (in GeV$^2$) for $\xi = 0$ (red dashed line), $\xi=1$ (blue dotted line), and $\xi = 3$ (black solid line). For $p^2 \lesssim -0.1$ GeV$^2$ the curves nearly lie on top of each other.  The lattice QCD data are taken from Refs. \cite{Bowman:2005vx} (red data points) and \cite{Oliveira:2018lln} (brown data points).   }
 \label{fig:MC&OGE}
\end{figure}

\begin{figure}[h]
 \center
\includegraphics[width=\linewidth]{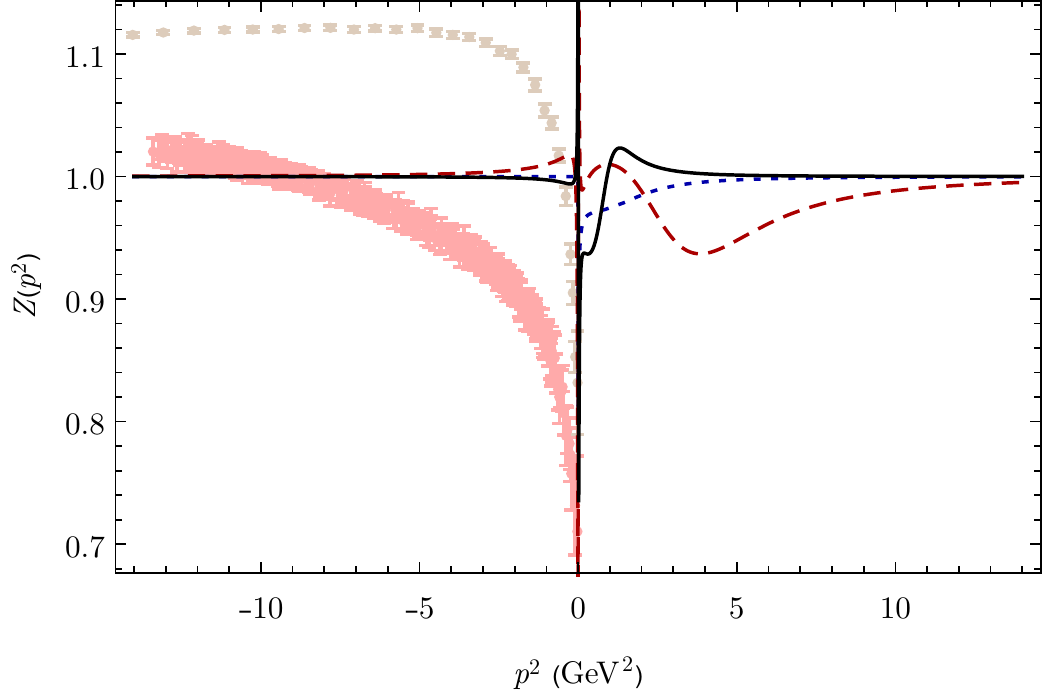}
\includegraphics[width=\linewidth]{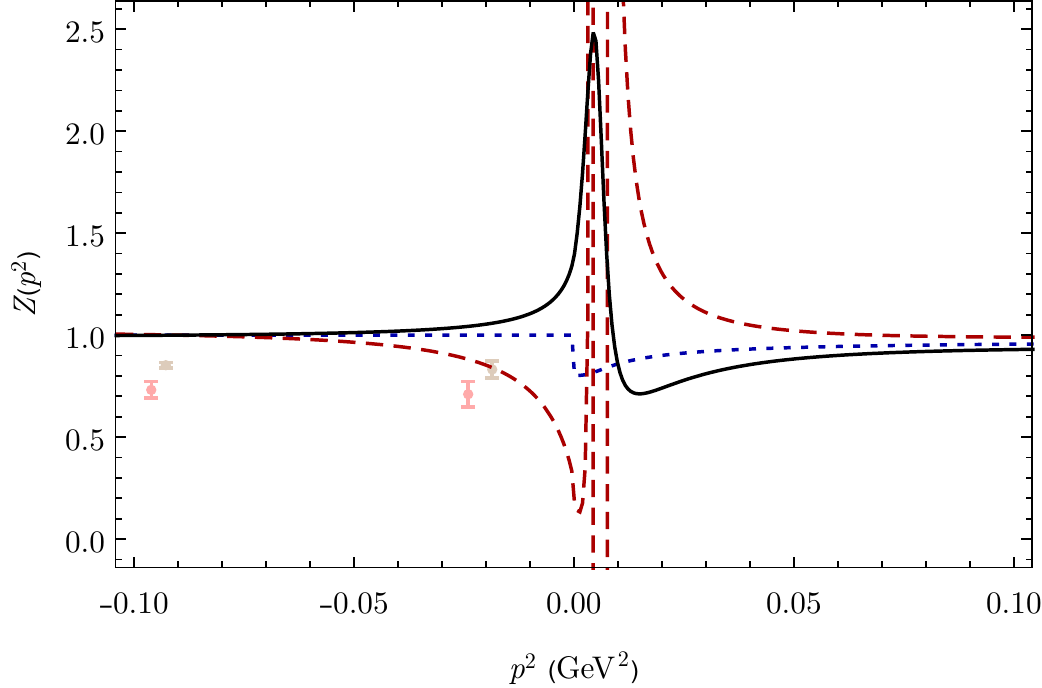}
 \caption{The wave function renormalization for the sum of constant and OGE kernels vs. $p^2$ (in GeV$^2$) for $\xi = 0$ (red dashed line), $\xi=1$ (blue dotted line), and $\xi = 3$ (black solid line). For $p^2 \lesssim -0.1$ GeV$^2$ the curves nearly lie on top of each other. The lattice QCD data are taken from Refs. \cite{Bowman:2005vx} (red data points) and \cite{Oliveira:2018lln} (brown data points).    }
  \label{fig:ZC&OGE}
\end{figure}

The results of this study are summarized in Figs.~\ref{fig:MC&OGE} and \ref{fig:ZC&OGE}.  They are sensitive to $\lambda_{\rm g}$,  and the parameters for the three cases are shown in Table \ref{tab:paras}. It should be stressed that only one parameter, $\lambda_{\rm g}$, has been roughly adjusted to agree with the data, while the other parameters were held fixed at some reasonable values given in Eq.~(\ref{eq:hgparas}) and the experimental strong coupling constant $\alpha_{\rm s}^{\rm p}=0.5$ (this choice, while consistent with our recent calculations of the heavy and heavy-light meson spectrum~\cite{Leitao:2017mlx,Leitao:2017it}, may be too small when more results are obtained for the light sector). The remaining parameters of Table \ref{tab:paras} are determined by the gap equation. The mass functions for the three gauges are indistinguishable in the spacelike region for $p^2\lesssim -0.1$ GeV$^2$ where they describe remarkably well the lattice data from Refs. \cite{Bowman:2005vx}. However, only the Yennie gauge ($\xi=3$) gives a $Z$ function that dips below zero for $p^2\le -8$ GeV$^2$ (but the effect is quite small), and this is another reason why we consider this gauge preferred over the others.

 \section{Summary and conclusions}
\label{sec:VI}

We have presented the first CST calculation of a quark mass function in Minkowski space, for both spacelike and timelike quark momenta, from the same quark interaction kernel that is used in calculations of the heavy- and heavy-light meson spectra. The calculations are performed in the chiral limit of vanishing bare quark mass. The kernel contains a OGE mechanism, a covariant generalization of a constant interaction, and a linear confining interaction~\cite{Leitao:2017mlx,Leitao:2017it}. However, we chose the confining interaction to be of purely Lorentz scalar and pseudoscalar type, such that the linear confining interaction does not contribute to the quark self-energy.

In previous work it was already shown how a dressed quark mass function in Minkowski space, that is consistent with D$\chi$SB, can be constructed in CST. However, only a simple constant interaction was used~\cite{Biernat:2014jt}. When employing a more realistic interaction kernel, including a OGE, this task becomes much more difficult, and one is initially faced with a number of new issues that require a careful treatment:

(i) A CST calculation of the quark self-energy that is consistent with the CST two-body calculations through chiral symmetry breaking requires the omission of the gluon propagator pole contributions in the loop integration (this study is limited to the use of a simple dressed gluon propagator with a constant mass). However, in this case the gluon poles can overlap with the quark poles and thus cannot be neglected anymore. This means that the basic idea behind the CST, namely that residues of poles in the kernel (OGE) are small compared to the ones of the quark propagators, breaks down. (ii) Another issue related to the omission of the gluon poles is due to divergent integrals of the self-energy invariants that lead to a pathological zero of the mass function near the origin, which is inconsistent with D$\chi$SB.  (iii) The gap equation for the constituent quark mass, when solved by taking the principal value of the gluon singularities, does not have a solution for positive $\alpha_{\rm s}$. 

We have shown in this work that issue (ii) can be dealt with by using a particular form factor that properly regularizes the integrals, while (i) and (iii) are resolved by introducing a gluon dressing, effectively giving the gluon a finite mass, and at the same time 
 implementing a prescription that makes the dressed gluon propagator non-singular (the same  ``Prescription C'' that has already been applied successfully in the CST theory of the NN and 3N systems \cite{Gross:2008ps}). The disadvantage of this method is that the mass function develops a discontinuity at  $p^2=0$. Fortunately, the size of the discontinuity depends on the gauge, which led us to study the behavior of our results in general linear covariant gauges, characterized by the continuous gauge parameter $\xi$. 

We find that the quark mass function is continuous  at $p^2=0$ only for one particular value of the gauge parameter, namely when $\xi=3$ (the so-called ``Yennie gauge''). This led us to elect the Yennie gauge as the ``the gauge of choice'' for our CST calculations. For comparison, we also present results obtained in the Landau ($\xi=0$) and in the  Feynman-'t Hooft gauge ($\xi=1$).
 For on-shell quantities, such as the constituent quark mass $m$, that may be considered an ``indirect observable'', or the on-shell  self-energy, we find that they are independent of the gauge -- a natural feature of the CST which is due to the decoupling of the $q^\mu q^\nu$ term of the kernel from the gap equation.

In the timelike region, except at the on-shell point, our mass functions depend quite strongly on the gauge, whereas in the spacelike region, except near $p^2=0$, the gauge dependence is very weak.
If one model parameter of our kernel is roughly adjusted, while the remaining parameters are given some reasonable values, our mass function in the spacelike region can also be brought into close agreement with the existing lattice QCD data. This can be seen as an indication that Prescription C for curing the problem of kernel singularities is working well.

For the wave function renormalization, $Z(p^2)$, our results exhibit a similar gauge-dependence as the mass function. They 
do not agree as closely with the lattice data as the mass function does. However, one can also observe a substantial variation between different sets of lattice data, such that no strong conclusions can be drawn from this comparison. Nevertheless,
preliminary studies with a running gluon mass suggest that the $p^2$ dependence of $Z(p^2)$ can still be modified.

The calculations of the dressed quark propagator in CST presented in this paper complete an important step towards our goal of constructing a covariant framework for the description of few-quark systems. It is now possible to use a realistic kernel together with consistently dressed quark propagators in a charge-conjugation-invariant CST calculation of bound states containing light quarks, and, in particular, to implement dynamical chiral symmetry breaking in pion systems.

}
\begin{acknowledgments}
We thank Gernot Eichmann for valuable discussions, in particular with respect to the gluon dressing functions. We thank Orlando Oliveira and Richard Williams for help with the lattice data. E.B., T.P. and A.S. also thank the Theory Group at Jefferson Lab for supporting several visits, during which part of this work was carried out. 
This work was funded in part by Funda\c c\~ao para a Ci\^encia e a 
Tecnologia (FCT) under Grants No. CFTP-FCT (UID/FIS/00777/2013), No. SFRH/BPD/100578/2014, and No. SFRH/BD/92637/2013. F.G. was supported by the U.S. Department of Energy, Office of Science, Office of Nuclear Physics under contract DE-AC05-06OR23177. Figure~\ref{fig:Vc} has been drawn using JaxoDraw~\cite{Binosi:2003yf}.
\end{acknowledgments}
\appendix

\section{Proof of gauge independence at $s=1$ }   
\label{app:gaugeindep}
In this appendix we prove that a $q^\mu q^\nu$-term of the interaction kernel considered in this paper does not contribute to the CST Dyson equation~(\ref{eq:DE}) at the on-shell point. As a consequence, the on-shell self-energy is gauge-parameter independent. Multiplying Eq.~(\ref{eq:DE}) with $[M^2(p^2)-p^2]$, taking the on-shell limit $p_0\to \sigma E_p$ and using the gap equation~(\ref{eq:oscondition}) yields
\bea
m+\slashed{\hat p}_\sigma&=&-\lim_{p_0\to \sigma E_p}S_0(p)Z_2\Sigma^\xi(\slashed{p})S(p)[M^2(p^2)-p^2]\,
\nonumber\\&=&-\lim_{p_0\to \sigma E_p}S_0(p)Z_2\left[\Sigma(\slashed{p})+(1-\xi)\Delta\Sigma(\slashed{p})\right]\nonumber\\&&\qquad\times\left[M(p^2)+\slashed p\right]\,,
\eea
where $\Sigma(\slashed{p})\equiv\Sigma^{\xi=1}(\slashed{p})$ and $(1-\xi)\Delta\Sigma(\slashed{p})$ are the self-energy contributions from the $\mathrm g^{\mu\nu}$- and the $q^\mu q^\nu$-terms of the kernel, respectively. Rewriting this equation gives
\bea
&&\left[-1-S_0(\hat p_\sigma)Z_2\Sigma(\slashed{\hat p}_\sigma)\right](m+\slashed{\hat p}_\sigma)\nonumber\\&&\;\;= 
\lim_{p_0\to \sigma E_p}S_0(p)(1-\xi)Z_2\Delta \Sigma(\slashed{p})\left[M(p^2)+\slashed p\right]
\nonumber\\
&&\;\;= \lim_{p_0\to \sigma E_p} (1-\xi)\sum_{\sigma'}\int_{\bf k} {\cal I}(\hat k_{\sigma'} ,p)\slashed{q}_{\sigma'} \,\Lambda({\hat k}_{\sigma'})\slashed{q}_{\sigma'} \nonumber\\&&\qquad\times\left[M(p^2)+\slashed p\right]
\nonumber\\
&&\;\;= 
\lim_{p_0\to \sigma E_p} (1-\xi)\sum_{\sigma'}\int_{\bf k} {\cal I}(\hat k_{\sigma'},p) \slashed{q}_{\sigma'}\Lambda({\hat k}_{\sigma'})\nonumber\\&&\qquad\times\left\{m M(p^2)-p^2+\left[m-M(p^2)\right]\slashed p \right\}= 0\label{eq:qqcontributionosprop}
\eea
where ${\cal I}(\hat k_{\sigma'} ,p)$ depends on the details of the kernel and must satisfy 
\bea \lim_{p_0\to \sigma E_p} q_{\sigma'}^2\, {\cal I}(\hat k_{\sigma'} ,p)=\mathrm{const.}
\eea
Equation~(\ref{eq:qqcontributionosprop}) is identical to the gap equation~(\ref{eq:oscondition}), as can be seen by multiplying~(\ref{eq:qqcontributionosprop}) with $S_{0}^{-1}(\hat p_\sigma)/2m$ from the left,
\bea
  (m-m_0)\Lambda ({\hat p}_\sigma)&=& Z_2\Sigma(\slashed{\hat p}_\sigma)\Lambda ({\hat p}_\sigma)\nonumber\\&=&m Z_2\left(\frac{A_0}{m}+B_0\right)\Lambda ({\hat p}_\sigma)\,.
   \eea
This shows that the $q^\mu q^\nu$-term of a Lorentz-vector interaction kernel, such as the OGE and constant kernels considered here, does not contribute to the CST Dyson equation at the on-shell point, and therefore does not contribute to the gap equation and the generation of the dressed quark mass $m$. 
\section{Reduction of the $\Sigma_{\rm g}$ integrals}   
\label{app:startnew}

Here we derive the results given in Sec.~\ref{sec:startnew}. For the analysis of the integrals (\ref{eq:Feynman}) and (\ref{eq:Rfun}), which were obtained in the rest frame where $p=p^{\rm r}\equiv \{p_0,{\bf0}\}$, it is convenient to scale out the quark mass $m$ by introducing the dimensionless variables $r_0=p_0/m$, $s=r_0^2$, $m_{\rm g}=M_{\rm g}/m$, and express the integrals in terms of the integration variable $y=E_k/m$.  Then
\bea
\frac{q_\sigma^2}{m^2}\equiv \rho_\sigma^2= 1+s-2\sigma y\,r_0\,, \qquad\frac{D_\sigma}{m^2}\equiv d_\sigma
\eea
and, after the angular integration, the $k$-integration becomes a $y$-integration,
\bea
\int_{\bf k}= \frac{m^2}{4\pi^2}\int_1^\infty \sqrt{y^2-1} \, \mathrm dy\, .
\eea
This gives the results 
\bea
\frac{\bar{A}_{\rm g}(s\,m^2)}{m} &=& \int_1^\infty \mathrm dy\sqrt{y^2-1}\,g(y) {\cal A}_{\rm g}(s,y)\,,\nonumber\\
\bar{B}_{\rm g}(s\,m^2)
&=&\int_1^\infty \mathrm dy\sqrt{y^2-1} \, g(y) {\cal B}_{\rm g}(s,y)\,,
\nonumber\\
\bar{R}_{\rm g}(s\,m^2)
&=&  \int_1^\infty \mathrm dy\sqrt{y^2-1}\,g(y){\cal R}_{\rm g}(s,y)\,,\qquad
\label{eq:321}
\eea
where
\bea
{\cal A}_{\rm g}(s,y) &=&\frac{8}{3\pi}\sum_{\sigma} \frac{1}{d_\sigma}\,,
\nonumber\\
{\cal B}_{\rm g}(s,y) &=&-\frac{4y}{3\pi\,r_0}\sum_\sigma\frac{\sigma}{d_\sigma}\,,
\nonumber\\
{\cal R}_{\rm g}(s,y) &=& \frac{4(y^2-1)}{3\pi} \sum_{\sigma} \frac{1}{\rho_\sigma^2 d_\sigma}\, .\qquad\quad
\label{eq:calABR}
\eea
\subsection{Timelike region $s>0$} 
To work out the implications of the absolute value of $q^2$ in Prescription C, we write $\rho_\sigma^2$ as
\bea
\rho_\sigma^2=2|r_0|\left(y_0 -\sigma y \frac{r_0}{|r_0|}\right)
\eea
where we recall that $y\ge1$ and
\bea
 y_0=\frac{1+s}{2|r_0|}\ge 1\, .
\eea
If $r_0 >0$, $\rho_-$ is always positive, but $\rho_+$ may be either positive or negative.  Conversely, if $r_0<0$, $\rho_+$ is always positive, but $\rho_-$ may be either positive or negative.  Because of the absolute values this separates the $y$ integration into two regions:  
\bea
{\rm if}\, y&\le& y_0:%
\quad  \rho_+^2>0,\, \rho_-^2 >0,\;{\rm for\; any}\;r_0
\quad\qquad{\rm Region}\, 1\,,
\nonumber\\
{\rm if}\, y&>& y_0:%
\quad \begin{cases} \rho_+^2<0,\, \rho_-^2 >0,\;{\rm for}\;r_0>0\cr
 \rho_+^2>0, \,\rho_-^2 <0,\;{\rm for}\;r_0<0
\end{cases}\quad{\rm Region}\, 2\, .
\nonumber\\
\eea
Notice that $y_0>1$ for all values of $r_0$, except $r_0=\pm1$, where $y_0=1$.  
\subsubsection{Region 1 ($y\le y_0$)}
For either sign of $r_0$ the denominators in Region 1 become
\begin{align}
d_+&\to m_{\rm g}^2+1+s-2y\,r_0 \,,\hfill &
\nonumber\\
d_-&\to m_{\rm g}^2+1+s+2y\,r_0\, .\hfill &
\end{align}

Hence the combination of factors for the functions in the integrands in Region 1 combine to give results which will lead to the removal of the factors linear in $r_0$.  Introducing the shorthand notation
\bea
\chi_1\equiv m_{\rm g}^2+1+s \label{eq:chi1}
\eea
we obtain the following results:
\bea
{\cal A}^1_{\rm g}(s,y)&=&%
\frac{16\chi_1}{3\pi(\chi_1^2-4y^2s)}\,,
\nonumber\\
{\cal B}^1_{\rm g}(s,y)&=&%
-\frac{16y^2}{3\pi(\chi_1^2-4y^2s)}\,,
\nonumber\\
{\cal R}_{\rm g}^1(s,y)&=&%
\frac{8(y^2-1)[(1+s)\chi_1+4y^2s]}{3\pi\rho_+^2\rho_-^2(\chi_1^2-4y^2s)} \, .\qquad
\label{eq:region11}
\eea
\subsubsection{Region 2 ($y> y_0$)}
In Region 2 the denominators depend on the sign of $r_0$:
\bea
{\rm if}\;r_0&>&0: \quad \begin{cases} 
d_+\to m_{\rm g}^2-1-s+2y\,r_0\cr
d_-\to m_{\rm g}^2+1+s+2y\,r_0\,,\end{cases}
\nonumber\\
{\rm if}\;r_0&<&0: \quad \begin{cases} 
d_+\to m_{\rm g}^2+1+s-2y\,r_0\cr
d_-\to m_{\rm g}^2-1-s-2y\,r_0\, .\end{cases} \label{eq:Din2}
\eea
Hence, the results depend on the sign of $r_0$, but the two cases in (\ref{eq:Din2}) can be combined if written in terms of $|r_0|$ instead of $r_0$.  Introducing the shorthand notation
\bea
\chi_2\equiv m_{\rm g}^2+2y\,|r_0|
\eea
the functions in the integrands %
become
\bea
{\cal A}^2_{\rm g}(s,y)&=&%
\frac{16\chi_2}{3\pi(\chi_2^2-(1+s)^2)}\,,
\nonumber\\
{\cal B}^2_{\rm g}(s,y)&=&
-\frac{8y(1+s)}{3\pi |r_0|(\chi_2^2-(1+s)^2)} \,,
\nonumber\\
{\cal R}_{\rm g}^2(s,y)&=&
\frac{8(y^2-1)(1+s)(\chi_2+2y\,|r_0|)}{3\pi \rho_+^2\rho_-^2(\chi_2^2-(1+s)^2)}\, .\qquad
\label{eq:region22}
\eea
These factors do not depend solely on $s=r_0^2$, but they are independent of the sign of $r_0$, and since they do not apply as $r_0\to0$ (notice that this point lies in Region 1, as discussed below), the apparent singularity in ${\cal B}_{\rm g}^2$ is never reached. 
\subsection{Spacelike region $s<0$}

For the calculation with $s<0$ we switch, for convenience, to a frame where $p=\tilde p^{\rm r}\equiv \{\mathrm i\,m\,r_0,{\bf 0}\}$ (with the choice $r_0>0$). The use of complex momenta is not a problem. The result is covariant and can therefore be transformed to real physical momenta by using a complex Lorentz transformation (see Appendix~\ref{app:slt0}). With this choice, the absolute value of the complex-valued function $\rho_\pm^2$ gives
\bea
|\rho_\pm^2|\equiv \rho_{\rm c}^2=\sqrt{(1+s)^2-4y^2s}
\eea
(where now $s=-r_0^2$).  Hence the functions in the integrands for $s<0$ (which we call Region 0) are
\bea
{\cal A}^0_{\rm g}(s,y)&=&%
\frac{16}{3\pi(m_{\rm g}^2+\rho_{\rm c}^2)}\,,
\nonumber\\
{\cal B}^0_{\rm g}(s,y)&=&%
0\,,
\nonumber\\
{\cal R}_{\rm g}^0(s,y)&=&%
\frac{8(y^2-1)(1+s)}{3\pi\rho_{\rm c}^4(m_{\rm g}^2+\rho_{\rm c}^2)}\, .  \label{eq:ABR01}
\eea
Notice that ${\cal R}_{\rm g}^0$ has a zero at $s=-1$.
\subsection{Limiting behavior and convergence of the integrals}
We first study the behavior of these results near $s=0$. When $s\to 0_+$, then $y_0 \to \infty$, thus the integrals are given by the results from Region 1, Eq.~(\ref{eq:region11}): 
\bea
{\cal A}_{\rm g}^1(0_+,y)&=&\frac{16}{3\pi (m_{\rm g}^2+1)}\,,
\nonumber\\
{\cal B}_{\rm g}^1(0_+,y)&=&-\frac{16y^2}{3\pi (m_{\rm g}^2+1)^2}\,,
\nonumber\\
{\cal R}_{\rm g}^1(0_+,y)&=&\frac{8(y^2-1)}{3\pi (m_{\rm g}^2+1)}\, . \label{eq:asy01}
\eea
However, as $s\to 0_-$ we must take the results from Region 0, Eq.~(\ref{eq:ABR01}):
\bea
{\cal A}_{\rm g}^0(0_-,y)&=&\frac{16}{3\pi (m_{\rm g}^2+1)}\,,
\nonumber\\
{\cal B}_{\rm g}^0(0_-,y)&=&0\,,
\nonumber\\
{\cal R}_{\rm g}^0(0_-,y)&=&\frac{8(y^2-1)}{3\pi (m_{\rm g}^2+1)}\, .
\eea
Notice that ${\cal B}_{\rm g}$ is not continuous at $s=0$. We can, however, obtain a $\bar{B}_{\rm g}^{\xi}(p_0^2)$, Eq.~(\ref{eq:AelBel}), that is continuous at $p_0^2=0$, if we  choose the gauge parameter as $\xi = 3$, which corresponds to the Yennie gauge~\cite{Fried:1958zz}.

Next, we study the behavior of the functions in the integrands for large $s$. When $s\to +\infty$, then $y_0\to +\infty$, and we again need only the results from Region 1:
\bea
{\cal A}^1_{\rm g}(s,y)&\stackrel{s\to \infty}{\longrightarrow}&\frac{16}{3\pi s}\,,
\nonumber\\
{\cal B}^1_{\rm g}(s,y)&\stackrel{s\to \infty}{\longrightarrow}&-\frac{16y^2}{3\pi s^2}\,,
\nonumber\\
{\cal R}_{\rm g}^1(s,y)&\stackrel{s\to \infty}{\longrightarrow}&\frac{8(y^2-1)}{3\pi s^2}\, . \label{eq:asy2}
\eea
For $s\to-\infty$ we get
\bea
{\cal A}^0_{\rm g}(s,y)&\stackrel{s\to -\infty}{\longrightarrow}&-\frac{16}{3\pi s}\,,
\nonumber\\
{\cal B}^0_{\rm g}(s,y)&=&0\,,
\nonumber\\
{\cal R}_{\rm g}^0(s,y)&\stackrel{s\to -\infty}{\longrightarrow}&-\frac{8(y^2-1)}{3\pi s^2}  . \label{eq:asy0}
\eea
We see that the asymptotic results for ${\cal A}_{\rm g}$ are symmetric and positive and the asymptotic behavior of ${\cal R}_{\rm g}$ is antisymmetric. 
The asymptotic behavior of ${\cal B}_{\rm g}$ shows, however, a problem similar to its behavior at $s=0$, which can also be fixed by choosing $\xi = 3$.  Note that the large $s$ behavior of the self-energy invariants is independent of the gluon mass, as expected.

\subsection{Prescription B}

For comparison, we also record here the results obtained with the Prescription B, {\it i.e.}, without using the Prescription C (see the discussion of Sec. \ref{sec:prescriptionC}). In that case the denominators are 
\bea
d_+&\to &m_{\rm g}^2-1-s+2y\,r_0\,
\nonumber\\
d_-&\to &m_{\rm g}^2-1-s-2y\,r_0\, ,
\eea
and the results (valid for all values of $s$) are
\bea
{\cal A}^{\rm B}_{\rm g}(s,y)&=&%
\frac{16\chi_{\rm B}}{3\pi(\chi_{\rm B}^2-4y^2s)}\,,
\nonumber\\
{\cal B}^{\rm B}_{\rm g}(s,y)&=&%
\frac{16y^2}{3\pi(\chi_{\rm B}^2-4y^2s)}\,,
\nonumber\\
{\cal R}_{\rm g}^{\rm B}(s,y)&=&%
\frac{8(y^2-1)[(1+s)\chi_{\rm B}-4y^2s]}{3\pi\rho_+^2\rho_-^2(\chi_{\rm B}^2-4y^2s)} \, ,\qquad
\label{eq:regionx}
\eea
where $\chi_{\rm B}=m_{\rm g}^2-1-s$.

{ \section{Complex Lorentz transformations}  \label{app:slt0}

Here we discuss the use of complex momenta in the calculation of the quark self-energy in the spacelike region where $p^2\le0$ ({\it i.e.} $s\le0$). The need for self-energy functions defined at $s\le0$ arises in the CST quark-quark scattering problem where the two quarks can have four-momenta  
\bea
\hat p_{1}&=&\{ E_{p},0,0, |{\bf p}|\}\,,
\nonumber\\
p_{2}&=&\{ E_{p}-\mu, 0,0, |{\bf p}|\}\,,  \label{eq:2bodymomenta}
\eea
where $P=\hat p_1-p_2=\{\mu,{\bf 0}\}$ is the momentum of the quark-quark system at rest, $\hat p_{1}$ is the on-shell momentum of quark 1 (in the $z$-direction for simplicity), and $p_{2}$ is the off-shell quark 2 with mass
\bea
p_{2}^2=\mu^2+m^2-2\mu E_{p}\equiv m^2 s\, .
\eea
When
\bea
{\bf p}^2\ge \frac{(\mu^2- m^2)^2}{4\mu^2}\,,
\eea
$s$ is negative and $m^2 s+{\bf p}^2$ is positive, and the four-momentum of quark 2 can be written
\bea
p_{2}&=&\{\sqrt{m^2s+{\bf p}^2},0,0,|{\bf p}|\}
\nonumber\\&=&\left\{\sqrt{-p_0^2+{\bf p}^2},0,0,|{\bf p}|\right\}\equiv  \tilde p'\, . \label{eq:p2}
\eea
Since the momenta (\ref{eq:2bodymomenta}) are real, it might seem appropriate to calculate the self-energy for $s<0$ in a standard frame where $p=\{0,0,0,p_0\}$, and obtain the result in the moving frame (\ref{eq:p2}) by a Lorentz transformation (LT).  In fact we have tried this and find that a whole new phenomenology is required in order to regulate the integrals, and this makes it difficult and somewhat arbitrary to relate the $s<0$ calculation to the $s>0$ one done in the rest frame where $p=p^{\rm r}\equiv \{p_0,{\bf 0}\} $.  Doing the calculation in the frame where $p=\tilde p^{\rm r}\equiv \{\mathrm i\,p_0,{\bf 0}\}$ is much more natural; the phenomenology required connects smoothly with that used for $s>0$.  

Since the self-energy calculation can be separated from the rest of the dynamics, we can use a complex LT to connect a calculation in the frame $p=\tilde p^{\mathrm r}$, \mbox{$m^2s=(\tilde p^{\rm r})^2=-p_0^2$}, to one in a moving frame with four-momenta (\ref{eq:p2}) (for a brief introduction to the complex Lorentz group, see {\it e.g.} Ref.~\cite{Streater:1989vi}). The transformation that accomplishes this, is given by 
\bea
 \tilde{\cal B}\left(|{\bf p}|\hat {\bf z}\right)=\frac{1}{\mathrm i p_0}\left(\begin{matrix} \sqrt{-p_0^2+{\bf p}^2} & 0 & 0 & |{\bf p}|\\[0.1in]
0 & \mathrm i p_0 & 0 & 0 \\[0.1in]
0 & 0 & \mathrm i p_0 & 0 \\[0.1in]
|{\bf p}| & 0 &  0 &  \sqrt{-p_0^2+{\bf p}^2} \end{matrix}\right)  \, \nonumber\\\label{eq:boost2}
\eea

such that
\bea
 \tilde{\cal B}\left(|{\bf p}|\hat {\bf z}\right) \tilde p^{\rm r}= \tilde p'\,, 
\eea 
where $\hat {\bf z}$ is the unit vector in the $z$-direction. To establish that this is a LT it is sufficient to show that it satisfies
\bea
\tilde{\cal B}^\intercal\left(|{\bf p}|\hat {\bf z}\right) G \tilde{\cal B}\left(|{\bf p}|\hat {\bf z}\right)=G\, ,
\eea
 where $G=\{{\mathrm g}^{\mu\nu}\}=\mathrm {diag}(1,-1,-1,-1)$.

Of course, equivalently we could have started with the four-momentum $ \tilde p'$ of (\ref{eq:p2}) and transformed it directly into $\tilde p^{\rm r}$ by means of the inverse of the transformation (\ref{eq:boost2}). What matters is that we can justify using the frame where $p=\tilde p^{\rm r}$ by the Lorentz invariance of our phenomenology.

}
\section{Proof of covariance of ${\cal V}_{\rm c}$}\label{app:covaVc}

The covariant constant kernel, defined in Eq.~(\ref{eq:Ckernel}), is used in this paper only when ${\bf p}=0$.  In this appendix we discuss how to use this kernel in general applications. First we focus on the case when $p^2 \ge 0$ and then generalize to results for $p^2<0$ by means of a complex LT (for the discussion of complex LT, see Appendix~\ref{app:slt0}).

\begin{figure}
 \centering
 \includegraphics[width=0.4 \linewidth]{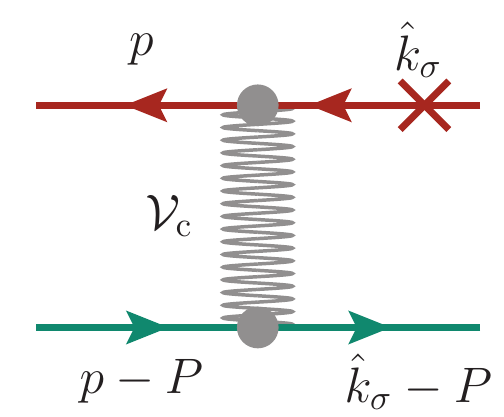}
 \caption{Diagrammatic representation of the constant kernel ${\cal V}_{\rm c}$ in the case where it is defined with an incoming quark with four-momentum $k=\hat k_\sigma=\{\sigma E_k,{\bf k}\}$ on-shell (denoted by the ${\bm \times}$ on the line).  In this case the quark may be either on its positive energy mass shell, with $k_0=E_k$, or on its negative energy mass shell with $k_0=-E_k$.}
 \label{fig:Vc}
\end{figure}

In the CST, the constant kernel is defined only when at least one quark is on-shell, as shown diagrammatically in Fig.~\ref{fig:Vc}.  
Here we assume that the incoming quark is on-shell with four-momentum $k=\hat k_\sigma=\{\sigma E_k,{\bf k}\}$ [either on its positive ($\sigma=+$) or negative energy ($\sigma=-$) shell].

\subsection{Timelike $p$}
 In this paper, we use the definition~(\ref{eq:Ckernel}) only when ${\bf p}=0$, {\it i.e.}, in the frame where $p=p^{\rm r}$. In this case~(\ref{eq:Ckernel}) becomes
\bea
{\cal V}_{\rm c}({ p^{\rm r},\hat k_\sigma})&=&\frac{CE_k}{2m}(2\pi)^3\delta^3\left({\bf k}\right) h(p_0^2)
\gamma_\mu\otimes\gamma_\nu \Delta_{\mathrm c}^{\mu\nu}(q_\sigma^2)\nonumber\\ \label{eq:Ckernel2}
\eea
where $q_\sigma^2=m^2+p_0^2-2\sigma E_k p_0 $. Because this kernel always acts under the integral $\int_{\bf k}$ it effectively replaces all $\hat k_\sigma$ with $\hat p^{\rm r}_{\sigma}\equiv \{\sigma m,{\bf 0}\}=\sigma \hat p^{\rm r}$ by the $\mathrm d^3{\bf k}$ integration. Here we show that the result in other frames can be obtained by boosting the ${\bf p}=0$ result to an arbitrary frame where ${\bf p}\ne 0$. Notice that~(\ref{eq:Ckernel2}) is not manifestly covariant, but nevertheless covariant because it is defined in a particular frame and can be generalized by a boost to an arbitrary frame, giving the expression (\ref{eq:Ckernel}). 

To show this we first consider the operator \bea
{\cal B}\left(|{\bf p}|\hat {\bf z}\right)=\frac{1}{p_0}\left(\begin{matrix} \sqrt{p_0^2+{\bf p}^2} & 0 & 0 & |{\bf p}|\\[0.1in]
0 & p_0 & 0 & 0 \\[0.1in]
0 & 0 & p_0 & 0 \\[0.1in]
|{\bf p}| & 0 &  0 &  \sqrt{p_0^2+{\bf p}^2} \end{matrix}\right)  \, ,\nonumber\\\label{eq:boost}
\eea
that boosts the four-vector $p^{\rm r}$ in the rest frame to \mbox{$p'=\{\sqrt{p_0^2+{\bf p}^2},0,0,|{\bf p}|\}$} in a moving frame in the $z$-direction. Notice that the complex LT of (\ref{eq:boost2}) is obtained from (\ref{eq:boost}) simply by replacing $p_0\to \mathrm i p_0$. The on-shell four-vector $\hat p_\sigma^{\rm r}$ transforms under the boost (\ref{eq:boost}) as
\bea
{\cal B}(|{\bf p}|\hat {\bf z}) \hat p_\sigma^{\rm r} \equiv \sigma \, \hat p'&=& \sigma \frac{m}{p_0} p'  \label{eq:pandpp}
\eea
and hence
\bea
\hat p'^2= \left(\frac{m}{p_0}\right)^2 p_0^2=m^2
\eea
as required by relativity. Similarly, a boost in an arbitrary direction $\hat {\bf p}$, denoted ${\cal B}({\bf p})$, which transforms $p^{\rm r}$ to $p'=\{\sqrt{p_0^2+{\bf p}^2},{\bf p}\}$, gives 
\bea
{\cal B}({\bf p}) \hat p_\sigma^{\rm r} = \sigma \hat p'&=& \sigma  \frac{m}{p_0} p' =\sigma\{E_{p'},{\bf p}'\}  \label{eq:pandpp1}
\eea with
\bea
{\bf p}'=\frac{m}{p_0} {\bf p}=\frac{m}{\sqrt{p^2}} {\bf p}\, , \label{eq:kandkp}
\eea
so the transformed $\sigma \hat p'$ is an on-shell four-vector, but with a three-vector part ${\bf p}'$ related to the three-vector part ${\bf p}$ of $p'$ by (\ref{eq:kandkp}). Therefore, the constant kernel in the boosted frame that provides  the replacement $\hat k_\sigma \to \sigma \hat p'$ under the integral $\int_{\bf k}$, is given by  
\bea
&&{\cal V}_{\rm c}({ p',\hat k_\sigma})\nonumber\\&&\;\;=\frac{C E_k}{2m}(2\pi)^3\delta^3\Big({\bf k}-\frac{m}{\sqrt{p'^2}}{\bf p}\Big) h(p'^2)
\gamma_\mu\otimes\gamma_\nu \Delta_{\mathrm c}^{\mu\nu}(q_\sigma^2) \nonumber\\&&\;\;=\frac{C E_k}{2m}(2\pi)^3\delta^3\Big({\bf k}-\frac{m}{p_0}{\bf p}\Big) h(p_0^2)
\gamma_\mu\otimes\gamma_\nu \Delta_{\mathrm c}^{\mu\nu}(q_\sigma^2)\,,\nonumber\\
\label{eq:Ckernel1}
\eea
where now
\bea
q_\sigma^2&=&m^2+p'^2-2\sigma E_k p'_0+2 {\bf k}\cdot {\bf p}\nonumber\\&=&m^2+p_0^2-2\sigma E_k \sqrt{p_0^2+{\bf p}^2}+2 {\bf k}\cdot {\bf p}\,.
\eea
\subsection{Spacelike $p$}
Next we consider the case of spacelike momenta $p$. The above expression~(\ref{eq:Ckernel1}) also holds for spacelike momenta, {\it i.e.}, when  $p'$ is replaced $\tilde p'$ with $\tilde p'^2<0$. Let $\tilde {\cal B}({\bf p})$ be a complex LT in arbitrary direction $\hat {\bf p}$ [defined similarly as (\ref{eq:boost2})] that transforms the timelike on-shell momentum $\hat p_\sigma^{\rm r}$ as
\bea
\tilde {\cal B}({\bf p}) \hat p_\sigma^{\rm r}  = \sigma \hat {\tilde p}'&=& \sigma  \frac{m}{\mathrm i p_0} \tilde p' =\sigma\{E_{\tilde p'},\tilde{\bf p}'\} \,, \label{eq:pandpptilde}
\eea 
with
\bea
\tilde {\bf p}'=\frac{m}{\mathrm i p_0} {\bf p}=\frac{m}{\sqrt{(\tilde p^{\rm r})^2}} {\bf p}\,  \label{eq:kandkptilde}
\eea
and 
\bea
\tilde p'=\left\{\sqrt{-p_0^2+{\bf p}^2},{\bf p} \right\}\,.
\eea
We find that the transformed momentum $\sigma \hat {\tilde p}'$ has complex three-vector components (\ref{eq:kandkptilde}), but is still is a timelike on-shell vector as required by relativity:
\bea
\hat {\tilde p}'^2=\left(\frac{m}{\mathrm i p_0}\right)^2 \tilde p'^2=\left(-\frac{m^2}{ p_0^2}\right) (- p_0^2)=m^2\,.\nonumber\\
\eea
Therefore, the covariant constant kernel ${\cal V}_{\rm c}(\tilde  p',\hat k_\sigma)$ with spacelike $\tilde p'$, obtained from transforming ${\cal V}_{\rm c}(\tilde  p^{\rm r},\hat k_\sigma)$ to arbitrary three-momenta ${\bf p}\neq 0$ by means of a complex LT $\tilde {\cal B}({\bf p})$, is given by
\bea
{\cal V}_{\rm c}({ \tilde p',\hat k_\sigma})&=&\frac{C E_k}{2m}(2\pi)^3\delta^3\Big({\bf k}-\frac{m}{\mathrm i p_0}{\bf p}\Big) h(-p_0^2)
\gamma_\mu\otimes\gamma_\nu \nonumber\\&&\quad\times\Delta_{\mathrm c}^{\mu\nu}(q_\sigma^2)\,,\label{eq:Ckernel1t}
\eea
where now
\bea
q_\sigma^2&=&m^2-p_0^2-2\sigma E_k \sqrt{-p_0^2+{\bf p}^2}+2 {\bf k}\cdot {\bf p}\,.
\eea
As anticipated, ${\cal V}_{\rm c}({ \tilde p',\hat k_\sigma})$ in~(\ref{eq:Ckernel1t}) is just the expression one obtains from~(\ref{eq:Ckernel1}) by replacing $p_0\to \mathrm i p_0$ and it provides under the  integral $\int_{\bf k}$ the replacement $\hat k_\sigma\to\sigma \hat {\tilde p}'$.
\\
\subsection{Quark self-energy}

Here we explicitly demonstrate the covariance of the quark self-energy calculated from ${\cal V}_{\rm c}$. In particular, we show that the function $A_{\rm c}^\xi$ obtained in Eq.~(\ref{eq:A_B_ell}) is unchanged by a boost. The self-energy in a frame with ${\bf p}\ne0$ is
\bea
\Sigma_{\rm c}^\xi(\slashed{p}')&=& \frac12\int_{\bf k}\sum_\sigma {\cal V}_{\rm c}(p',\hat k_\sigma) \Lambda({\hat k}_\sigma)\nonumber\\&=&
\frac{Z_2 C h(p'^2)}{8m}\sum_\sigma\gamma^\mu\left(m+\sigma\slashed{\hat p}'\right)\gamma^\nu\Delta_{\mathrm c}^{\mu\nu}(q_\sigma^2)\nonumber\\&=&
\frac{Z_2 C h(p'^2)}{8m}\sum_\sigma\Bigg\{ \gamma^\mu\left(m+\sigma\slashed{\hat p}'\right) \gamma_\mu\nonumber\\&&\quad-\left(1-\xi\right)\frac{\slashed q_\sigma \left(m+\sigma\slashed{\hat p}'\right)\slashed q_\sigma }{q_\sigma^2}\Bigg\}\nonumber\\&=&
 \frac{Z_2 Ch(p'^2)}{8m}\Bigg\{ 2m\left(3+\xi\right)\nonumber\\&&\quad-\sum_\sigma\sigma\left[2\slashed{\hat p}' +\left(1-\xi\right)\frac{\slashed q_\sigma \slashed{\hat p}'\slashed q_\sigma }{q_\sigma^2}\right]\Bigg\}\nonumber\\&=& 
 \frac14\left[3+\xi\right]Z_2 C h(p'^2)
=A_{\rm c}^\xi(p'^2)\, , 
\label{eq:Sigmac3}
\eea

where we have used in the last step that 

\bea
\frac{\slashed q_\sigma \slashed{\hat p}'\slashed q_\sigma }{q_\sigma^2}=
[{\cal B}^{-1} ({\bf p}) \gamma]^0\slashed{\hat p}'[{\cal B}^{-1} ({\bf p}) \gamma]^0 
\eea
is independent of $\sigma$ because 
\bea
\slashed q_\sigma&=&[\sigma \hat p'-p']_\mu \gamma^\mu= [{\cal B} ({\bf p}) (\sigma\hat p^{\rm r}-p^{\rm r})]_\mu \gamma^\mu\nonumber\\&=&  (\sigma m-p_0) [{\cal B}^{-1} ({\bf p}) \gamma]^0\,,
\eea
and $q_\sigma^2=(\sigma m-p_0)^2$.
Therefore, the correct result is recovered, which coincides with (\ref{eq:A_B_ell}) because $p_0^2=p'^2$. Analogously, for spacelike momenta $\tilde p'$ we obtain
\bea
\Sigma_{\rm c}^\xi(\slashed{\tilde p}')&=&  
 \frac14\left[3+\xi\right]Z_2 C h(\tilde p'^2)
=A_{\rm c}^\xi(\tilde p'^2)=A_{\rm c}^\xi(-p_0^2)\, , \nonumber\\
\label{eq:Sigmac4}
\eea
where we have used in the calculation
\bea
\slashed q_\sigma&=& (\sigma m-\mathrm i p_0) [\tilde {\cal B}^{-1} ({\bf p}) \gamma]^0\,
\eea
and $q_\sigma^2=(\sigma m-\mathrm ip_0)^2$. The result (\ref{eq:Sigmac4}) is obtained from (\ref{eq:A_B_ell}) by replacing with $p_0\to \mathrm i p_0$, as anticipated.

\bibliographystyle{h-physrev3}
\bibliography{PapersDB-v2-3}

\end{document}